\documentclass[apj]{emulateapj}
\usepackage{natbib}

\pdfoutput=1
\usepackage{ graphicx }
\usepackage{ latexsym }
\usepackage{ amsmath, amssymb }        
\usepackage{multirow}
\usepackage{longtable}
\usepackage{tabu}
\usepackage{aas_macros}
\usepackage[normalem]{ulem}

\newcommand{\Mesa}{\textsc{Mesa}}
\newcommand{\Enzo}{\textsc{Enzo}}
\newcommand{\YT}{\textsc{yt}}
\newcommand{\Celmo}{\textsc{Celmo}}
\newcommand{\eff}{\mathrm{eff}}

\newcommand{\msun}{M$_\odot$}
\newcommand{\lsun}{L$_\odot$}
\newcommand{\rsun}{R$_\odot$}
\newcommand{\Rsun}{\mathrm{R}_\odot}
\newcommand{\Lsun}{\mathrm{L}_\odot}

\newcommand{\kms}{km~s$^{-1}$}

\newcommand{\rd}{\mathrm{d}}
\newcommand{\h}{\mathrm{h}}
\newcommand{\e}{\mathrm{e}}

\newcommand{\bfa}{\textbf{(a)}}
\newcommand{\bfb}{\textbf{(b)}}
\newcommand{\bfc}{\textbf{(c)}}
\newcommand{\bfd}{\textbf{(d)}}

\def\lessim{\raise-.5ex\hbox{$\buildrel<\over{\scriptstyle\mathtt{\sim}}$}}
\def\grtsim{\raise-.5ex\hbox{$\buildrel>\over{\scriptstyle\mathtt{\sim}}$}}

\usepackage{color}

\begin{document}

\shorttitle{Common envelope light-curve -- I.}
\title{Common envelope light-curves -- I. grid-code module calibration}

\author{Pablo Galaviz\altaffilmark{1,2}, Orsola De Marco\altaffilmark{1,2}, Jean-Claude Passy \altaffilmark{3,*}, Jan E. Staff \altaffilmark{1,2} \and Roberto Iaconi \altaffilmark{1,2}}
\email{Pablo.Galaviz@me.com}
\altaffiltext{1}{Department   of    Physics   and   Astronomy,   Macquarie University, Sydney, NSW, Australia}
\altaffiltext{2}{Astronomy, Astrophysics and Astrophotonics Research Centre,  Macquarie University, Sydney, NSW, Australia}
\altaffiltext{3}{Argelander-Institut  f\"ur Astronomie, Auf dem H\"ugel 71, D-53121 Bonn, Germany}
\altaffiltext{*}{Alexander-von-Humboldt fellow}



\begin{abstract} 

  The common envelope binary  interaction occurs when a star transfers
  mass onto a  companion that cannot fully accrete it.   The interaction can
  lead to a merger of the two objects or to a close binary. The common
  envelope interaction is the gateway of all evolved compact binaries,
  all  stellar  mergers and  likely  many  of  the stellar  transients
  witnessed  to  date.   Common  envelope simulations  are  needed  to
  understand  this interaction  and  to interpret  stars and  binaries
  thought to be the byproduct  of this stage. At this time, simulations
  are  unable to reproduce  the few  observational data  available and
  several   ideas   have   been   put   forward   to   address   their
  shortcomings. The need for  more definitive simulation validation is
  pressing,  and   is already being  fulfilled  by   observations  from
  time-domain surveys.  In this  article, we present an initial method
  and   its  implementation   for   post-processing  grid-based common   envelope
  simulations to produce the  light-curve so as to compare simulations
  with  upcoming observations. Here  we implemented  a zeroth  order  method to
  calculate  the  light  emitted  from  common  envelope  hydrodynamic
  simulations carried out  with the 3D hydrodynamic code  Enzo used in
  unigrid mode.   The code implements an approach  for the computation
  of luminosity in both optically thick and optically thin regimes and
  is tested using the first  135 days of the common envelope simulation
  of  \citet{Passy2012},  where  a  0.8~\msun\ red  giant  branch  star
  interacts  with  a  0.6~\msun\ companion.   This  code  is  used  to
  highlight  two  large obstacles  that  need  to  be overcome  before
  realistic light curves  can be calculated. We explain  the nature of
  these  problems and  the attempted  solutions and  approximations in
  full  detail  to   enable  the  next  step  to   be  identified  and
  implemented. We also discuss our simulation in relation to recent data of transients identified as common envelope interactions.

\end{abstract}




\maketitle




\section{Introduction} 
\label{sec:introduction}


The common envelope (CE) interaction  between two stars has become the
standard explanation for the  existence of close evolved binaries such
as  cataclysmic variables  or the  progenitors of  Type  Ia supernovae
\citep{IvaJusChen13}.   Yet,  this interaction  continues  to elude  a
reasonable physical  description. Without it, it  becomes difficult to
carry     out      meaningful     population     synthesis     studies
\citep[e.g.,][]{Politano2007,Politano2010,Dominik2012}, including those allowing us to reconcile predicted and observed rates of gravitational-wave producing events \citep[e.g.,][]{Dominik2012,Dominik2015}.    Hydrodynamic
models   have    been   carried   out   with   a    range   of   codes
\citep{Rasio1996,Sandquist1998,Ricker2012,Passy2012,Nandez2014,Ohlmann2016,Iaconi2017}, but it  appears that even the  basics of the interaction,  such as the
final separation  or how much and  when the CE is  ejected, are poorly
reproduced by these models. 

Comparing model outputs with observations has mainly been limited to
post-CE  systems \citep[e.g.,][]{Schreiber2003,Zorotovic2010,DeMarco2011}.  The
separations of post-CE  systems tend to be larger  in some simulations than
in observations \citep{DeMarco2011,Passy2012,Iaconi2017}, although it is clear that simulated primaries with more massive and/or more compact envelopes result in smaller orbital separations \citep{Ohlmann2016,Iaconi2017}.  

The assumption is that post-CE systems  are generated  by the entire  removal of  the stellar envelope over one dynamic event. Most CE simulation do not succeed in
ejecting the entire envelope \citep{Sandquist1998,Ricker2012,Passy2012,Staff2016,Iaconi2017}. Recently, some simulations have successfully achieved envelope ejection by assuming that the entire recombination energy budget is available for the ejection \citep{Nandez2015,Nandez2016}. Even so, successful ejection only takes place for certain parameters \citep{Nandez2016}.  
Moreover, some recombination energy may escape, as the neutral medium becomes optically thin \citep{Harpaz1998}. As  a result of the discrepancies  between simulations and
observations  it  is non-trivial  to  use  the observations  as  code
validation (for a review of the CE problem see \citealt{IvaJusChen13}), as one may suspect additional physics or phases, not modelled in the simulations, may play a role  \citep[e.g.,][]{Kuruwita2016,Ivanova2016}.  

Recently,  time-resolved  observations have  detected  a  range of  new
outbursts,  which  have  been  named intermediate  luminosity  optical
transients (ILOT; \citealt{Kasliwal2012,Kashi2013}), so called because
they  have  intermediate luminosities   between  those  of  novae  and
supernovae and lead to extremely red outburst products.  Some of these
ILOTs  appear  to have  been  due  to the  merger  of  two stars.   In
particular the V1309~Sco ILOT \citep{Tylenda2011} has almost certainly
been caused by the merger of a subgiant and a low
mass  companion,  because the  contact  binary  was actually  observed
before  the outburst,  the period  was reducing,  and  the post-merger
object, a large giant, shows  no sign of binarity.  Other such objects
may have been V838~Mon \citep{Bond2003}, V4332~Sgr \citep{Martini1999}
as well as other, extragalactic  ones such as M31~RV \citep{Mould1990}
or M85~OT2007 \citep{Kulkarni2007} or M31~LRN2015 \citep{MacLeod2016}.  These outbursts, may  give us an
early glimpse  into the light properties of CEs and  hence provide  us with
additional model constraints  and code validation.  

As data accumulates we are already glimpsing at the complexity of these phenomena. It is clear that there are various phases characterising these presumed mergers: a phase preceding the dynamical merger, the dynamical merger itself,  and a phase following it, all of which have distinct light properties that contribute to the overall light behaviour \citep{MacLeod2016}. The possible processes that change a slowly evolving binary to a fast merging one are several, including the Darwin-instability \citep{MacLeod2016} or a slower merger driven by mass loss through the outer Lagrangian point \citep{Pejcha2016,Pejcha2016b}. As observations and models multiply, the role of CE simulation becomes a less and less isolated one and different codes and methods will have to be merged, or at least laid alongside \citep[for a review of the range of codes applied to these problems see][]{DeMarco2017}.        

In so doing CE codes will have to evolve to their next generation, with higher resolution \citep{Ohlmann2016} and the addition of extra physics, such as a more refined equation of state \citep{Nandez2015} or the addition of magnetic fields \citep{Ohlmann2016b}. In particular, radiation hydrodynamics will be a fundamental component  in understanding the light expected from the CE fast in-spiral phase. This step will allow us to understand when a CE takes place and when other emission systems are dominating the light.

In this paper we attempt the calculation of the light properties of one  simulation of the CE early fast-in-spiral phase by post-processing  one of the  hydrodynamic simulations  of  \citet{Passy2012}, hereafter P12.   The challenges presented by the CE binary interaction when attempting to extract the light properties from simulations are even greater than those encountered when tying to determine the gas dynamics. 
However these challenges need to  be quantified  in order  to improve the
calculation to the point of being useful. Quantifying the challenges to the accurate calculation of a CE light curve can also focus future hydrodynamic efforts towards aspect of the computation that can aid the post-processing of the light.

The paper is organised  as follows: in Section~\ref{sec:phys-situation} we describe the physical situation of the early in-spiral of the CE interaction between a giant and a less massive companion. In so doing we set the stage and introduce some of the challenges. In Section~\ref{sec:lumin-calc} we summarise  the  luminosity  calculation  approach, with details left to the appendix, where we emphasize the challenge of knowing the photospheric temperature.  This  is followed  by the  calculation of  the lightcurve  for  the CE simulations presented by P12 in Section~\ref{sec:results}. We then discuss available observational constraints in Section~\ref{sec:observations}. Conclusions      and       discussions are      presented in Section~\ref{sec:discussion}.

\section{The physical situation} \label{sec:phys-situation}

Before attempting the calculation of the light, it is important to define the physical regime and the parameters of the calculation. The simulation we base this work on is {\it Enzo2} of P12, carried out between a 0.88~\msun, 89~\rsun, RGB star and a 0.6~\msun, point mass companion. Their simulation was carried out using a domain size of 2~AU and 128 cells on a side\footnote{$Enzo2$ was not the most resolved simulation of P12, but the outcome of this simulation were not too different from their $Enzo7$ simulation, which had twice the resolution.}. Here we have repeated the simulation using the same code and setup, but with a domain four times as large and a 512 cells on a side so as to maintain the resolution identical (the cell size is 3.4~\rsun). The reason for this was to prolong the time during which the CE gas remains in the computational domain. 

\begin{figure*}
  \centering
  \includegraphics[width=146mm,clip=true,trim=-6mm 12mm 0mm 0mm]{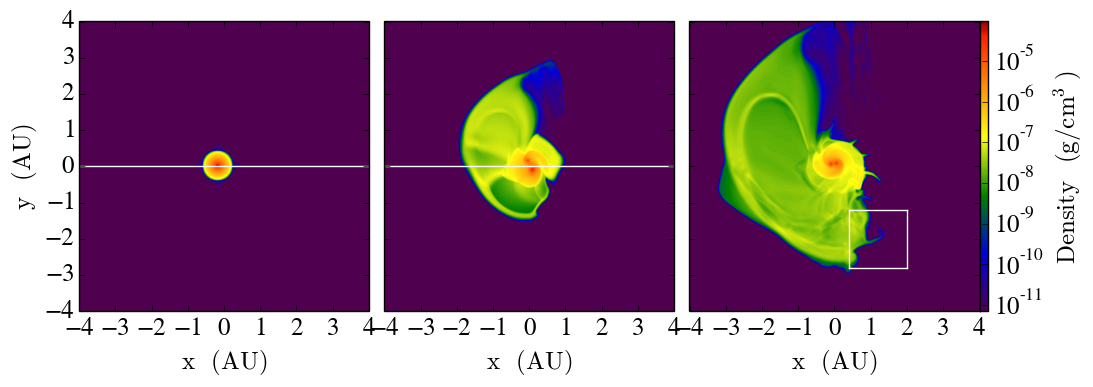}
    \includegraphics[width=140mm,clip=true,trim=0mm 83mm 0mm 85mm]{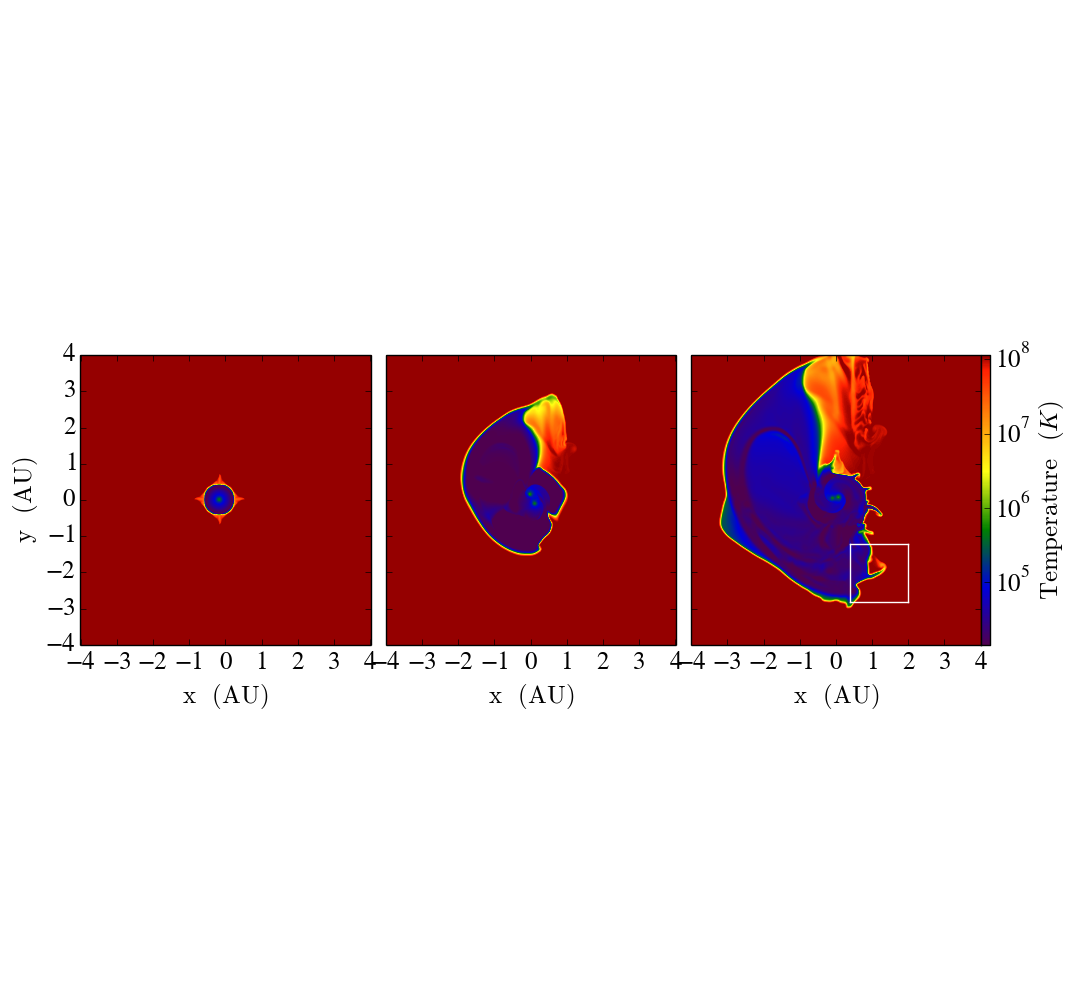}
    \includegraphics[width=140mm,clip=true,trim=0mm 83mm 0mm 85mm]{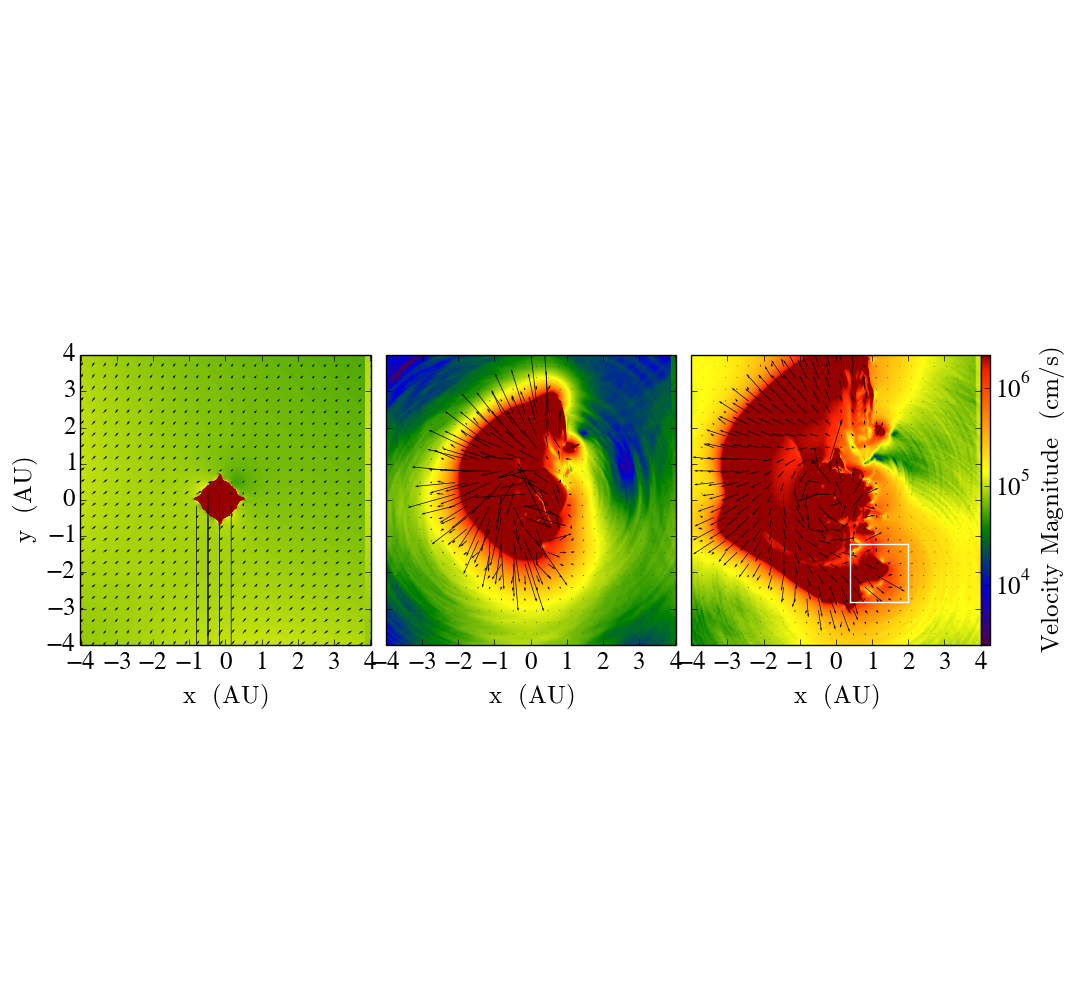}  
    \includegraphics[width=140mm,clip=true,trim=0cm 83mm 0mm 85mm]{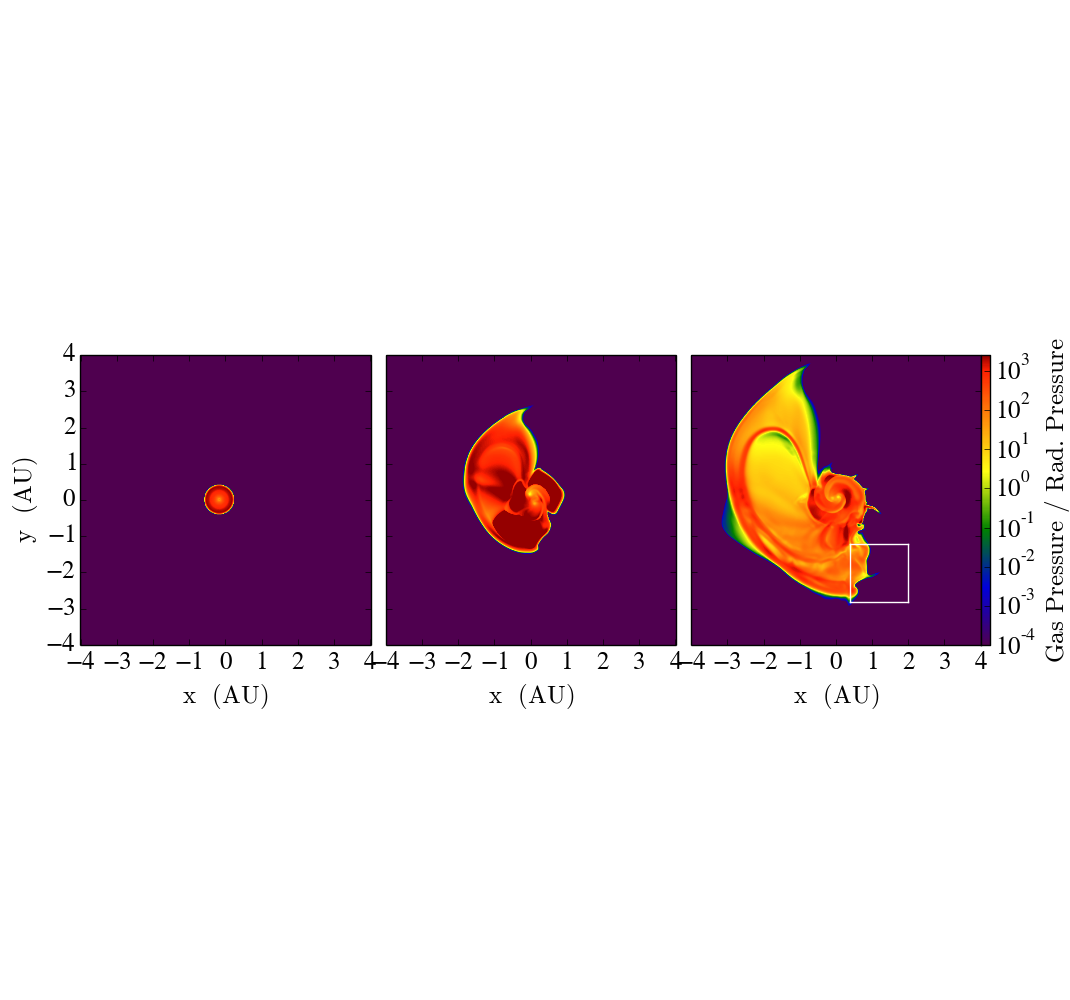}
        \includegraphics[width=140mm,clip=true,trim=0mm 70mm 0mm 85mm]{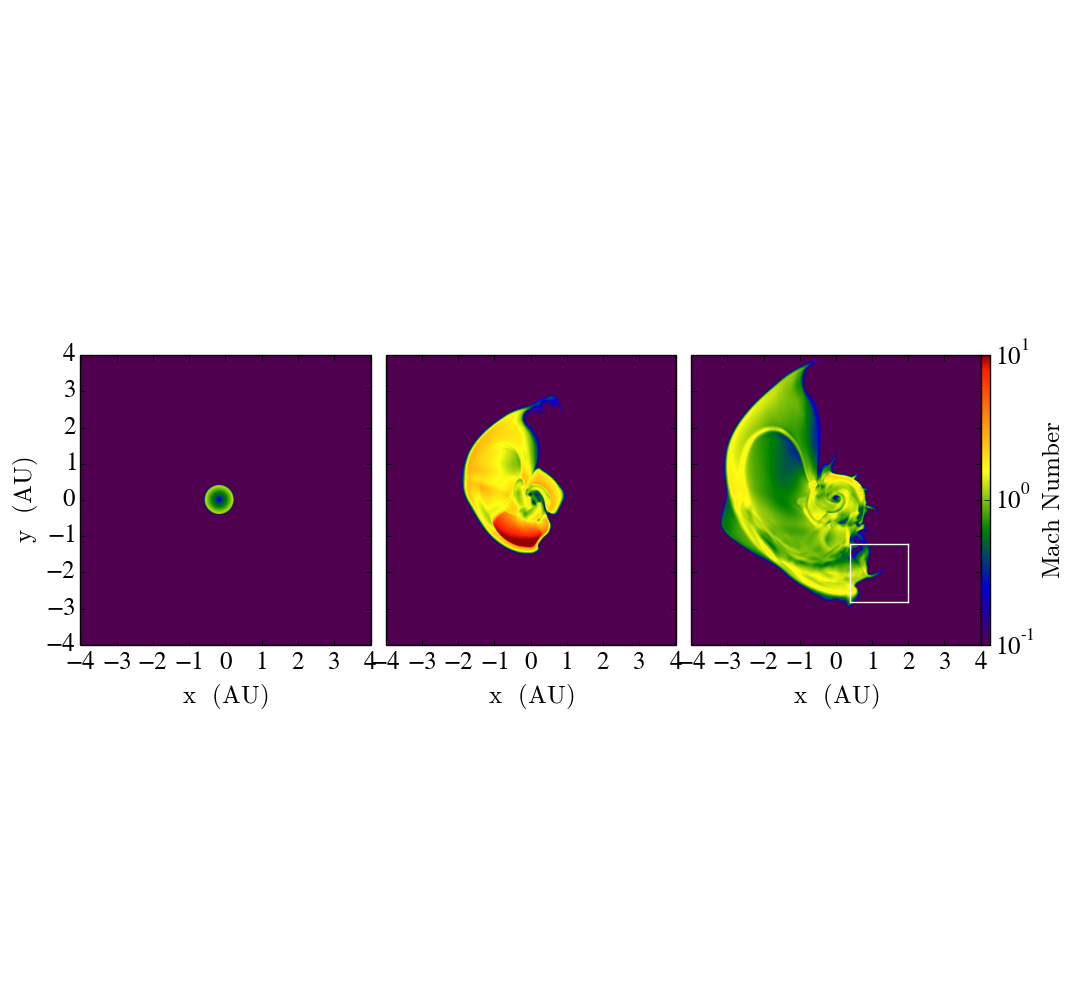}
     \caption{ Slices on the equatorial plane of density (row 1), temperature (row 2), velocity of the gas
  (row 3), ratio between the gas pressure and the radiation pressure (row 4), and Mach number (row 5). The slices are for $t=0$ (left column), 
  $t=75$ (middle column), and $t=135$~days (right column). The arrows show the direction 
  of the velocity and are normalised to the maximum value. In the vacuum, the ratio between the gas pressure and the radiation pressure is of the order $10^{-13}$.
  The white box delimits the close-up region presented in Fig.~\ref{fig:slices_detail}. Horizontal white lines mark the direction of 1-dimensional cuts presented in future figures.}
  \label{fig:slices}
\end{figure*}

\begin{figure*}
  \centering
 \includegraphics[width=150mm,trim={0.5cm 4.5cm 0.5cm 4cm},clip]{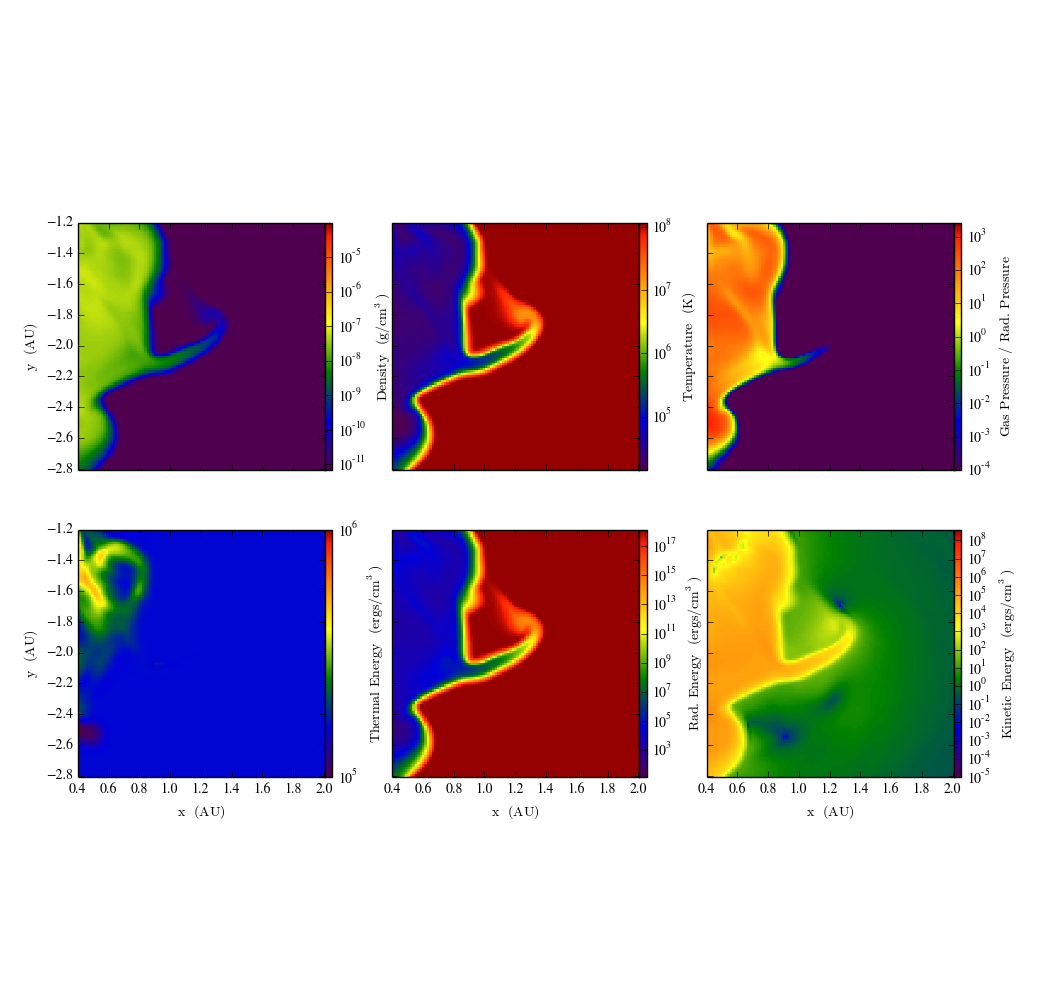}
  \caption{Details of slices on the equatorial plane of density (top left), temperature (top middle), gas pressure divided by radiation pressure (top right), thermal energy density (bottom left), radiation energy density (bottom middle) and kinetic energy density (bottom right), at $t=135$~days from the beginning of the simulation. The centre of each panel is approximately at $x=1.2$~AU and $y=-2$~AU from the centre of the domain (see Fig.~\ref{fig:slices}).}\label{fig:slices_detail}
\end{figure*}

In Fig.~\ref{fig:slices} we show a slice along the equatorial plane at three times during the simulation, at the beginning, right after the one-dimensional (1D) stellar structure has been mapped and stabilised, at 75 days and at 135 days. We display density, temperature, velocity modulus, the ratio of gas to radiation pressure and the Mach number. In Fig.~\ref{fig:slices_detail} we show a zoomed in detail of some of these quantities. Here we can also see the discrete nature of the grid and its relatively low resolution. 

\begin{figure*}
  \centering
  \includegraphics[width=100mm]{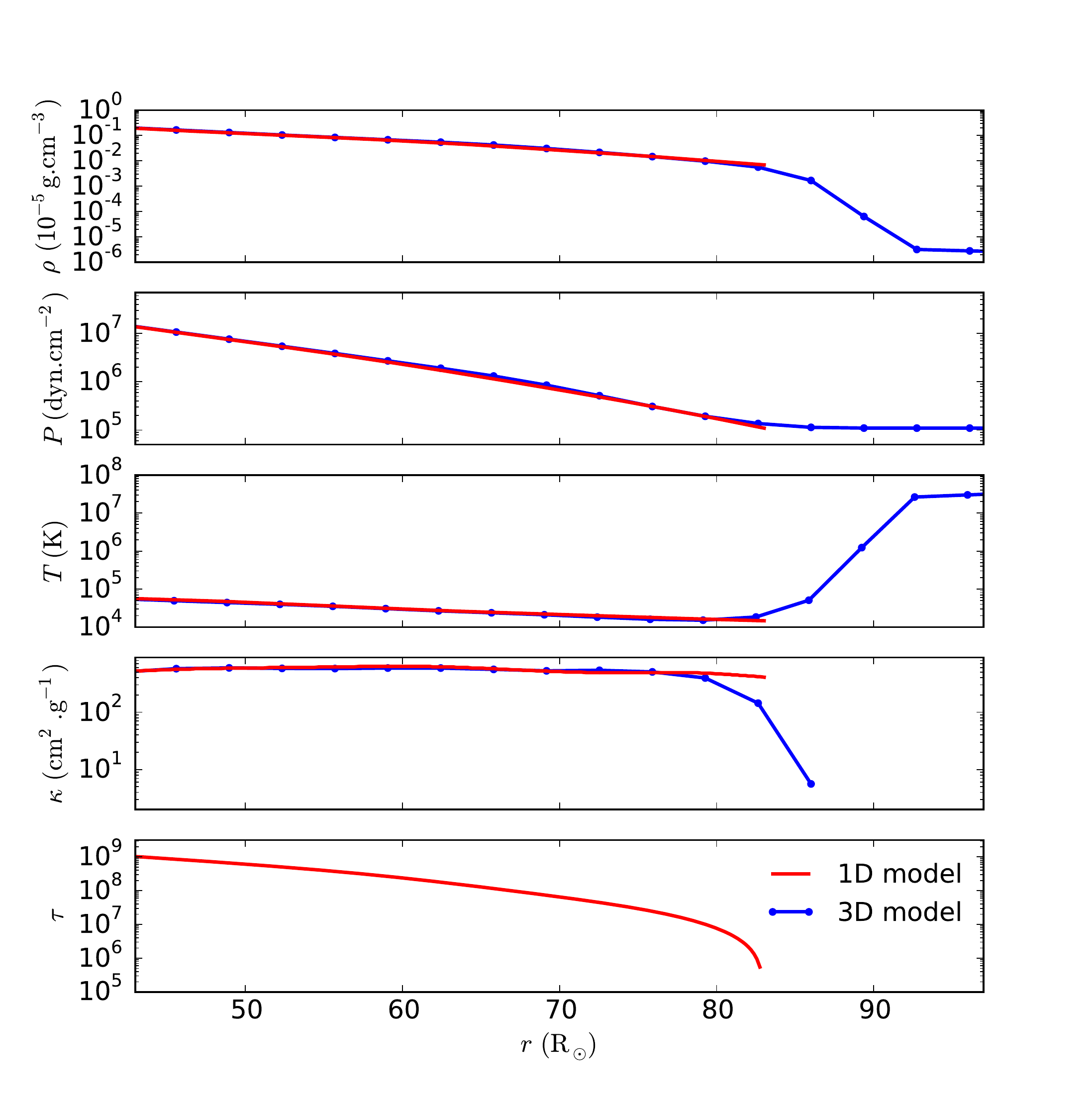}
  \caption{1D  (red lines) vs 3D (blue lines) comparison: the radial profile in 3D is  a ray in the $z$ direction which contains the origin.   The different panels compare the interpolations of the
    density (row 1), the gas pressure (row 2),  temperature (row 3),
    opacity (row 4), and optical depth (row 5).}
    \label{fig:3}
\end{figure*}

In Fig.~\ref{fig:3} we show 1D cuts at time zero where we can see the problem of mapping a higher resolution, 1D stellar structure onto a three-dimensional (3D) computational domain with far inferior resolution. The photosphere, as defined by the 1D model, has values of pressure, temperature, density, etc. which are vastly different from those encountered near the centre of the star. These changes are well captured by the 1D model, but lost as soon as it is interpolated onto the 3D domain. This is why in Fig.~\ref{fig:3} even the 1D model (red curves) is missing the points associated with the cooler, low density,  photospheric layers: they are all contained within one cell of the 3D domain.

\begin{figure}
  \centering
  \includegraphics[width=90mm]{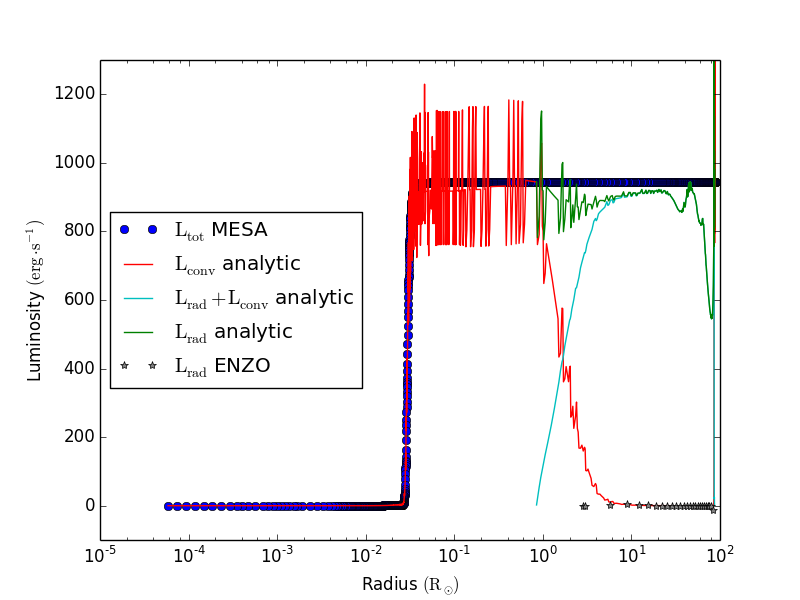}
  \caption{ A comparison of the stellar luminosity as a function of radius calculated in 1D (blue symbols), compared with that derived from the analytical expression of Eq.~\ref{eq:0} (red curve) and that calculated using the analytical expression after the star was mapped in 3D (green stars). The convective flux and luminosity was also calculated analytically (cyan line) and added to the radiative luminosity (red line) to make up total luminosity (green line).}
\label{fig:4}
\end{figure}

In Fig.~\ref{fig:4} we show an important aspect of the convective giant star that will become important at a later time, when we face the problem of the photospheric temperature. The luminosity of each layer in a radiative, spherical star is always equal to:  

\begin{equation}
L(r) = - 4 \pi r^2 {c \over 3 \kappa(r) \rho(r)} {\partial u_{rad}(r) \over \partial r}, 
\label{eq:0}
\end{equation}

\noindent where $r$ is the radius, $c$ the speed of light, $\kappa$ the opacity, $\rho$ the density and $u_{\rm rad}$ is the radiative energy density. This is not so for the convective layers where it is the bulk motion of the convective eddies or plumes that transports out the energy. Hence, applying the above expression to our 1D model, we see how in the deep convective envelope the radiative luminosity is small, while it increases to the total value in the outer thin radiative layer.


In the 3D simulation the gas is adiabatic and this must be a reasonable approximation 
because of the short duration of the expansion. The only heating is at the hand of compression and some shock heating early in the simulation (Fig.~\ref{fig:slices}). Some CE interactions do result in stronger shocks, but not all. In the interaction simulated by P12, the 0.6~\msun\ companion moves subsonically (Iaconi et al. 2017) although we do see locally mildly supersonic gas before 135 days. 

The high temperature of the gas around the star (see the temperature panels in Fig.~\ref{fig:slices}) is an artificial expedient commonly used in this type of grid computations \citep[e.g.,][]{Sandquist1998}. It insures that the stellar surface does not expand into the vacuum by providing a pressure that balances the atmospheric pressure by way of a very high temperature, but very low density ``vacuum gas". While this expedient has no consequence for the hydrodynamics, it is very problematic when extracting the light properties of the CE. As the simulation progresses, the outermost layers of the CE, which have lower densities, acquire a relatively large temperature as they ``mix" with vacuum gas. These layers are dynamically unimportant, but they have an artificially high temperature and high opacity. These thin, hot layers can be seen clearly in the temperature panels in Fig.~\ref{fig:slices} and Fig.~\ref{fig:slices_detail}, even at time zero, where a yellow ``skin" surrounds the the gas distribution.

Like these artificially hot layers, the low density ``vacuum" is completely opaque. 
Any model that attempts to calculate the light from these simulations will have to devise a way to avoid the low density ``vacuum" as well as any low density gas which has an unrealistically high temperature. As we explain below, we do this by imposing a ``density floor": gas with density lower than this floor is completely ignored in the calculation of the optical depth. In the detail in Fig.~\ref{fig:slices_detail}, top row, we see a small, low density, high temperature plume that is eliminated by the density floor.

In certain physical situations, such as supernova explosions, the relationship between gas thermal and radiation energies and expansion is such that photons can leak out from behind the photosphere. This is not the case here.  A CE expansion during the dynamical in-spiral is a relatively slow process more akin to the expansion of a Mira giant during its radial pulsation cycle than the expansion of an envelope during a supernova eruption. 

The early expansion phase which we are trying to characterise here is extremely optically thick, with almost none of the expanding gas becoming transparent. The speed at which the expansion takes place is of the order of tens of kilometres per second. In Fig.~\ref{fig:slices}, third row, we see that the expansion early in the interaction is below 50~km~s$^{-1}$ with only few pockets of material moving faster (the maximum velocity witnessed in the first 135 days of the simulation is 200~\kms). The Mach number of the gas is just over unity by 135 days and decreasing, with the exception of a pocket of gas at 75 days, which eventually disappears. Over the entire simulation, gas pressure dominates over radiation pressure, except in the ``vacuum" and within the thin skin of gas binding the envelope, that is heated by the external medium. The timescale of in-spiral is of the order of a year, with the expansion continuing beyond this time frame with a decreasing velocity.

\begin{figure}
  \centering
  \includegraphics[width=80mm]{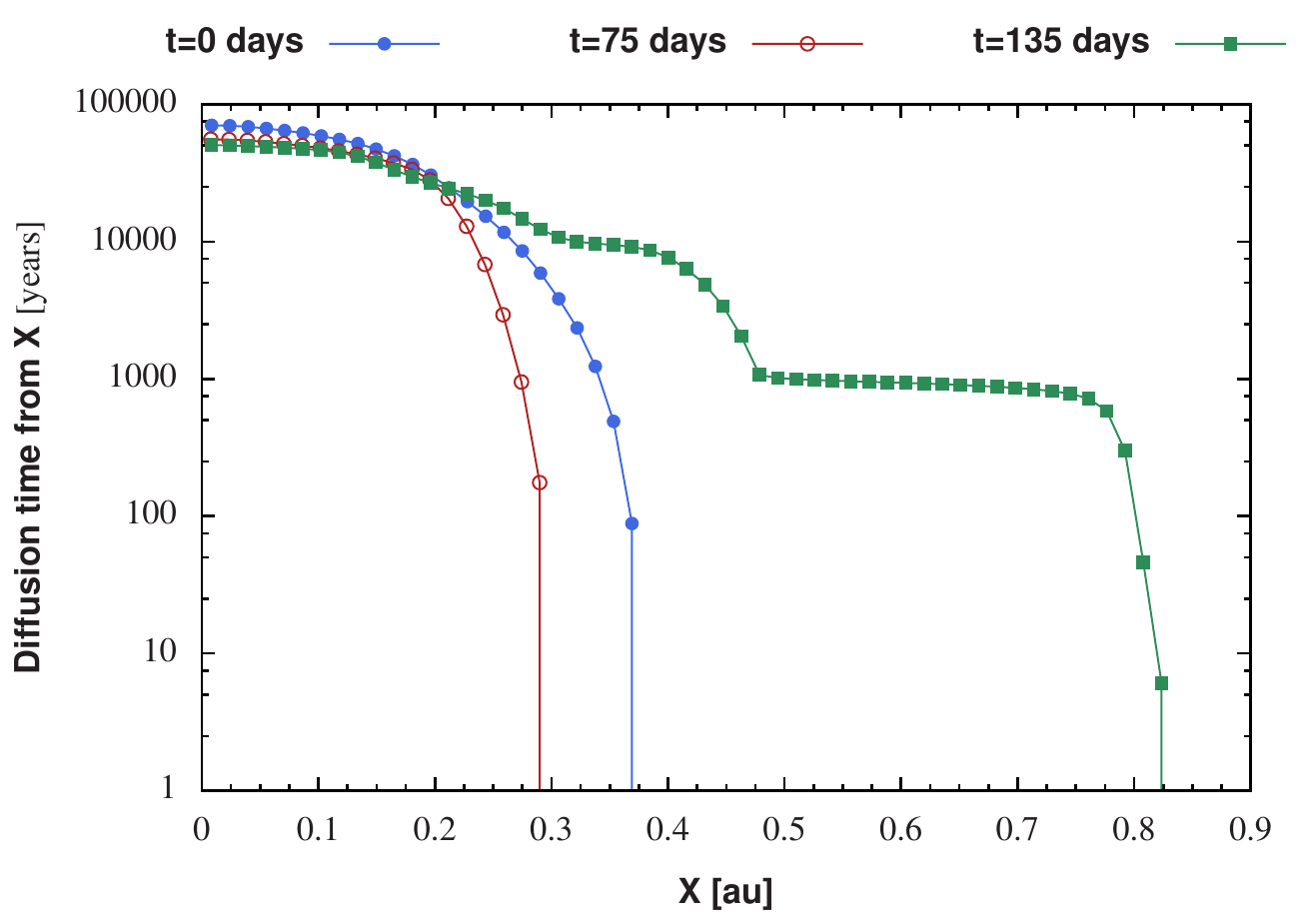}
\caption{Diffusion time taken by a photon emitted at a distance $x$ from the centre to reach the photosphere. 
  The calculations are made for the gas configurations observed at $t=0$~(blue), 75 (red), and 135 (green) days.}
  \label{fig:diff_times}
\end{figure}

Additionally, none of the energy associated with the in-spiral escapes during the short time of the in-spiral.  As the companion in-spirals the gravitational energy is deposited into the gas in the form, primarily, of thermal energy and to a much lesser degree of kinetic energy of the orbit as well as of the gas itself. The thermal energy will escape the star but on timescales longer than 135 days. In Fig.~\ref{fig:diff_times} we show the time photons would take to travel from a certain depth in the CE to the photosphere. This calculation is carried out by using a simple random walk theory with unequal step sizes \citep{Mitalas1992}, and is demonstrated for different times during the CE interaction (0, 75 and 135 days). As we can see the time that the photons would take to travel out is always longer than the time over which we want to calculate the light-curve, namely 135 days. Hence in the first 135 days we do not expect any of the energy deposited into the CE by the in-spiralling companion to escape.

\begin{figure}
  \centering
  \includegraphics[width=80mm]{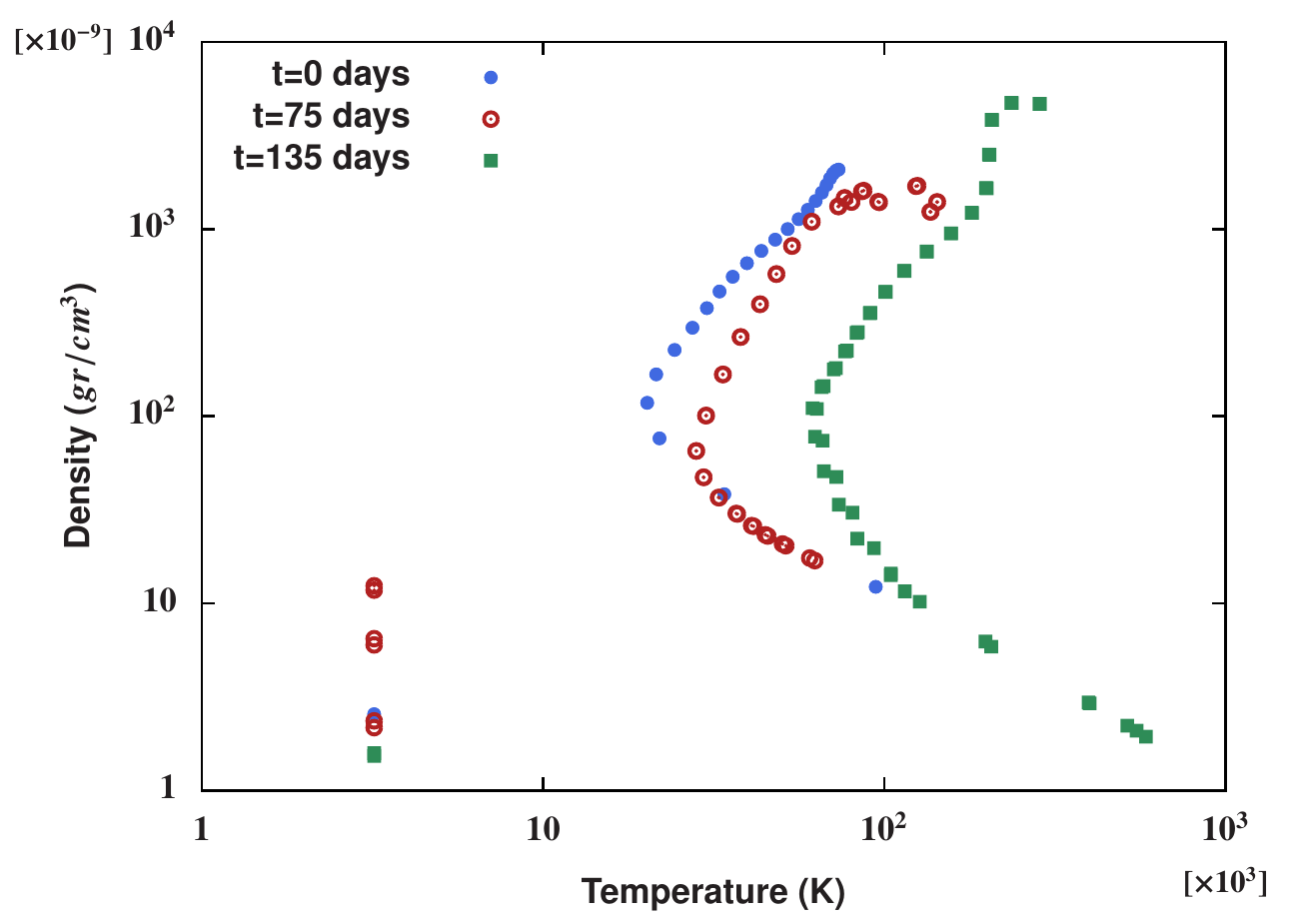}
\caption{Density-temperature profiles at three times during the simulation.}
  \label{fig:rho_T}
\end{figure}

In Fig.~\ref{fig:rho_T} we display the values of temperature and density along a ray from the centre of the domain to the domain boundary along the positive $x$ direction, at the usual three times during the simulation. The data points in the lower left corner of the plot, at low density and temperature are those located outside the photosphere, within the hot vacuum that is excluded in our simulation. Here we can see that all cells considered contain ionised gas (T$>$10\,000~K), so the dominating opacity source is from electron scattering. 

\section{Luminosity calculation and the temperature problem} \label{sec:lumin-calc}

The CE Light MOdule (\Celmo) reads the density and internal energy of
each volume element in the  3D computational domain for
each  time  step  for  which  the hydrodynamic  code  has  created  an
output. From  the internal  energy, a  temperature is  determined and,
using the  density and temperature,  an opacity is  interpolated using
opacity tables.  In opacity tables the opacities are expressed as  a function of $\log T$ and $\log
R$, where $R = \log \rho - 3 \log T + 18$. We have used opacity tables
for    Z=0.02     and    X=0.7    from     \citet{Grevesse1998}    and
\citet{Alexander1994}  with  metals  from  \citet{Grevesse1993}.   The
latter  table  extends  to  temperatures  as low  as  1000~K  (in  the
hydrodynamic simulations used in this  paper, the temperature is never
below this value). 
Using the density  and the opacity, the optical depth
is integrated  for each  volume element along  parallel rays  that are
perpendicular to  each face of the  numerical domain. In this  way the
location of the  surface where the optical depth is  2/3, is found and
the temperature  of that location used (but see Section~\ref{sec:temperature}),  assuming blackbody radiation,
to determine the brightness of each volume element.
In Appendix~\ref{sec:formulation} we describe these steps in detail. In Appendix \ref{ssec:filter} we describe the convolution with filter bandpasses, while in Appendix~\ref{sec:verification} we perform numerical tests to verify our implementation.

\subsection{The calculation of temperature} \label{sec:temperature}


When the 1D model is  mapped into the \Enzo\ computational domain,
the temperature quantity is not  required, since the specific internal energy of  each cell, $u_{\rm  int}$, is calculated from  the pressure
and density.   The \Enzo~ equation  of state is  that of an  ideal gas
with an adiabatic index $\gamma=5/3$, while the 1D star was calculated
with a  more sophisticated,  depth-dependent equation of  state.  When
the  1D star  is  mapped into  the  3D  domain it  is  not in  perfect
hydrostatic equilibrium.  This is why,  after the initial mapping, the
star   needs   to   be   stabilised  in   \Enzo,   as   described   in
P12.   The new  equilibrium model  tends to  be slightly
larger than the  original 1D model. 

This slightly  larger star constitutes  the initial  model in  \Enzo, which we  use for  the
initial  calculation  of  the luminosity.   Fig.~\ref{fig:3}  shows  a
typical comparison between the the 1D model and the one used in the 3D model after relaxation, where we are zooming in onto the outer part of the star. Here the 1D model is missing the data points characterising the photosphere. These data points are all at almost the same radius, have very low density, contain almost no mass, but can create a problem when the star is immersed in the hot vacuum. The photosphere is therefore usually eliminated from the 1D model at the time of mapping it into 3D. Even if we had retained the 1D photosphere, the contact with the hot vacuum would, by the second time step, have heated these layers to an unrealistically high temperature, generating the problem which we discuss further below.

The temperature at each cell centre in the 3D code is given by:
\begin{equation} T  =
\frac{M}{\overline R}(\gamma-1)u_{\mathrm{int}}, \label{eq:14}
\end{equation}
\noindent where  $\overline R$ is the universal gas  constant, $M=\mu m_H N_A$ is the molar
mass, $\mu$ is the mean molecular weight, $m_H$ is the hydrogen atomic
mass and $N_A$ is Avogadro's number. The temperature changes depending
on the composition.   The photospheric temperature for the  simulation presented in
Section~\ref{sec:results}  is smaller  than  10\,000~K,  which is  the
approximate limit for a neutral gas.  We therefore choose to use a mean
molecular weight of  1.26, corresponding to neutral  mass fractions of
$X=0.73$, $Y=0.25$ and $Z=0.02$.
\begin{figure*}
  \centering
\includegraphics[width=85mm,clip=true,trim=-6mm 0mm 15mm 0mm]{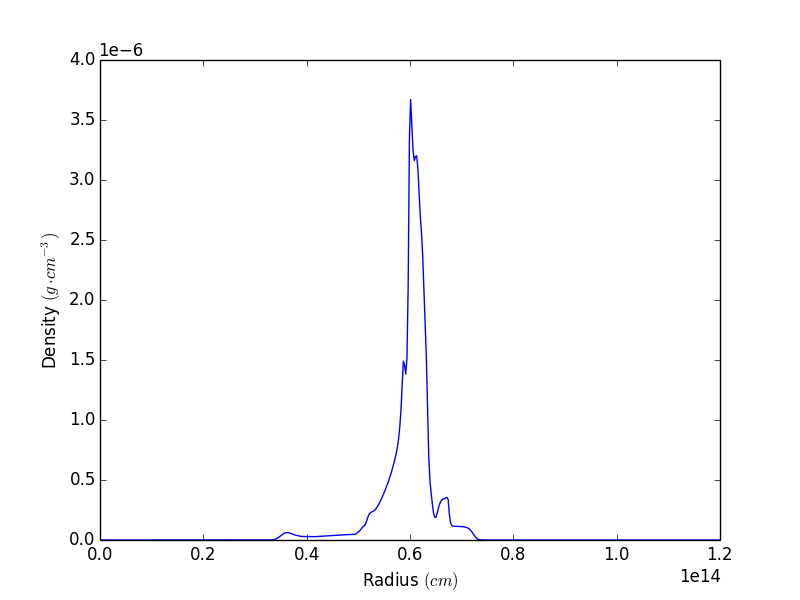}
\includegraphics[width=80mm,clip=true,trim=7mm 0mm 15mm 0mm]{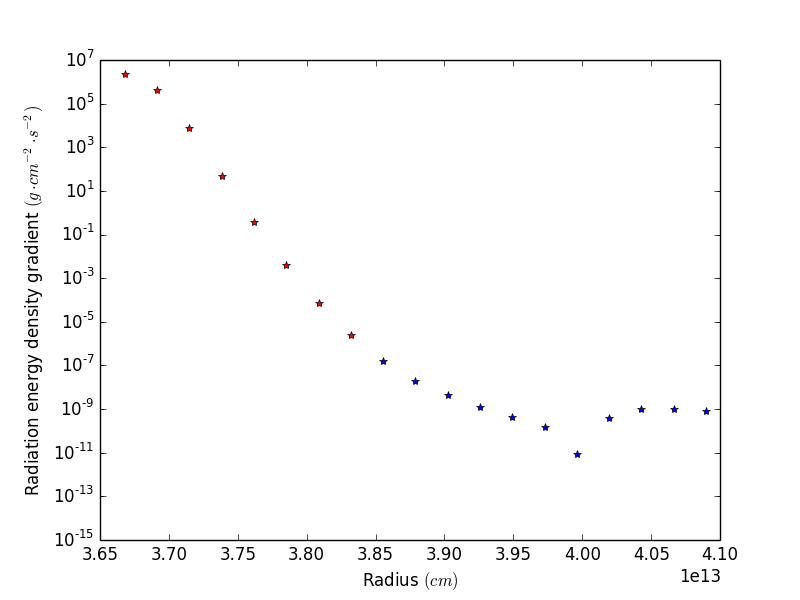}
     \caption{Left: A density cut at 75 days along the x-axis at $y$=$z$=4~AU (see ray plotted in Fig.~\ref{fig:slices}). Right: the values of $\partial u_{\rm rad} / \partial r$ at cell boundaries, centred on the left hand side of the density distribution at the location of the photosphere (blue symbols are inside the photosphere, while red symbols are outside).}
  \label{fig:delu-delr}
\end{figure*}

\subsection{The problem of spatial resolution in defining the effective temperature}
\label{ssec:resolutionproblem}

The optical depth $\tau$, determines  the amount of radiation which is
visible to an observer.  The photosphere is located at a surface where
$\tau=2/3$  (approximately half  of  the radiation  is visible).   The
volume of fluid above the photosphere defines the optical thickness of
the  fluid.   We  define  two  regimes in  our  simulation.   We  call
optically  thick  every ray  for  which the  back  of  the first  cell
occupied  by  gas  with   density  above  the  \Celmo\  density  floor
(first discussed at in Sec.~\ref{sec:phys-situation} and better described in Sec.~\ref{sssec:vacuum-temperature}),  has  an  optical depth  larger
than 2/3.

For the optically thick fluid that characterise the early 
expansion of the CE, the
$\tau=2/3$ surface is located to the precision of the hydrodynamic code resolution. 
The greater problem
is that the gradient of temperature  within the cell that contains the photosphere is very
steep. Consequently, while the physical location of the photosphere is
affected  by a  modest uncertainty,  the estimation  of the  effective
temperature, and hence of the luminosity is far more inaccurate.

Within  the cell  that  contains the  photosphere,  the optical  depth
ranges between zero at the ``front" side  of the cell, to a value much
larger than  unity, at the  ``back'' side. The temperature  at the centre of the
same cell  can have  an arbitrarily  high value  because of  the steep
temperature gradient  in the proximity  of the stellar  photosphere. A
straight interpolation  between the  two cells in front and behind the
cell containing  the photosphere,  is meaningless.  The cell  in front
usually has a temperature value  related to the ``vacuum'' temperature
discussed      in      Sec.~\ref{sec:phys-situation},      hence
unrealistically high. The temperature at the centre of the cell  ``behind" the cell containing the photosphere has the  high value  characteristic of a  location farther  inside the
star.   Hence   finding  the   correct  photospheric   temperature  is
impossible  by interpolation,  because  we have  no  knowledge of  the
external value.  Using  a value of zero, or similar  low value for the
cell just  outside that containing  the photosphere, and using  one or
even multiple points  to interpolate the temperature at  centre of the
cell containing  the photosphere,  gives a  range of  possible values,
which are dependent on arbitrary fit parameters. These fitting methods 
give an answer for the temperature to within a factor of 2, but the $T^4$ 
dependance of the luminosity makes these uncertainties unacceptable.

Below we discuss a second problem inherent to grid-based CE simulations that has an even worse impact on our attempt to calculate the light from the interaction.

\subsection{The ``vacuum" temperature problem}  
\label{sssec:vacuum-temperature}  
  
As explained in Sec.~\ref{sec:phys-situation}, in grid simulations the vacuum outside the 
star  cannot be empty, because otherwise the 
star diffuses rapidly out \citep[e.g., see][P12]{Sandquist1998}.  To obviate this problem
the vacuum is replaced with a  very low density medium. The density is
a factor of $10^{-4}$ smaller than the lowest stellar density. In
the case of the  simulations presented in Section~\ref{sec:results} the \Enzo\
density floor is
$7 \times  10^{-12}$~g~cm$^{-3}$.  The low  density medium is  kept in
pressure  equilibrium   with  the  star  surface  by   having  a  high
temperature ($\sim 10^8$~K). 

Such high ``vacuum" temperature would be optically  thick, so  \Celmo~ has  a minimum  density below  which the medium is considered completely  optically thin. This ``density floor"
has to be  larger than the density floor in  the \Enzo\ simulation for
the  following reason.  At the beginning  of  the  CE  simulation low density, 
hydrodynamically unimportant 
``fingers" of stellar gas expand into  the vacuum and their temperature
is affected by the high ``vacuum"
temperature, so  these low density features have unrealistically high temperature 
and are therefore optically thick, artificially extending the photosphere. This problem disappears
rapidly as  more mass expands  and dynamically overwhelms  the tenuous
vacuum medium. To circumvent this  problem we keep the \Celmo~ density
floor at $5 \times 10^{-10}$~g~cm$^{-3}$. This density floor would affect the computation of the light in later phases of the expansion, where the medium has expanded sufficiently to decrease in density. However, for the optically thick, early part of the interaction considered here the exact value of the density floor has no effect on the determination of the photospheric location.

A far greater problem is that the hot vacuum warms up any outer stellar layers that are adjacent to it and which have lower density. This can clearly be seen by comparing the density and temperature panels both in Figs.~\ref{fig:slices} and \ref{fig:slices_detail}. These outer layers are those where we seek to extract the value of the temperature and we see that even if the resolution were higher, their temperature is compromised by the hot vacuum.

\subsection{Effective temperature determination from flux conservation} 
\label{sec:optic-thick-fluid}

An alternative approach is calculating the luminosity using the radiative flux across a thin layer located behind the overheated photosphere. In Fig.~\ref{fig:delu-delr}, left panel, we show a density cut at 75 days, along a line marked in Fig.~\ref{fig:slices} (top row, middle panel). Along this line we read values of the density, opacity, and we calculated values of $\partial u_{\rm rad} / \partial r$. Values of $\partial u_{\rm rad} / \partial r$ across the photosphere on the left hand side of the gas distribution are plotted in Fig.~\ref{fig:delu-delr}, right panel. The red symbols in Fig.~\ref{fig:delu-delr}, right panel, are values of the gradient characterising the hot vacuum just to the left of the photosphere, while the blue symbols are values just inside the photosphere. The symbols are 3.4~\rsun\ apart, the resolution of the grid. As can be seen, the cells straddling the photosphere have quite a range of values of $\partial u_{\rm rad} / \partial r$. Just inside the photosphere the first $\sim$8-10 cells are affected by artificial heating as can be seen in the temperature panel in Fig.~\ref{fig:slices_detail}. Values of the gradient in the cells immediately behind those are $\sim$10$^{-9}$~erg~cm$^{-4}$. Values of the opacity at corresponding depth is $\sim$100~cm$^2$~g$^{-1}$, while the density is $\sim$$4\times 10^{-8}$~g~cm$^{-3}$.  

In order to check on the viability of this scheme we assume that the distribution of gas is spherical and adopt Eq.~\ref{eq:0} with a mean radius value calculated as the radius of the sphere that has the same volume as that contained by the photosphere at that time. This is 180~\rsun\ (see Fig.~\ref{fig:radius}). With this radius we calculate an approximate luminosity of $\sim$1~\lsun, much lower than even the initial stellar luminosity of 648~\lsun, while we expect a value similar or larger. Aside from the uncertainty affecting the choice of the right values for gradient and opacity, the reason for this discrepancy must be related to the thermodynamic properties of the gas. At the location we sampled, the quantities reflect the  convective envelope where the flux is not transported by radiation, similar to what shown in Fig.~\ref{fig:4} for time zero. This exercise is unlikely to provide us with a meaningful value of the overall luminosity of the gas distribution even if, instead of assuming spherical symmetry we calculated the flux at that pixel, assumed that it characterises the photosphere and integrated to obtain the luminosity using the actual shape of the gas distribution.

\subsection{Effective temperature determination from a stratified temperature distribution}
\begin{figure*}
  \centering
\includegraphics[width=160mm,clip=true,trim=0mm 0mm 0mm 0mm]{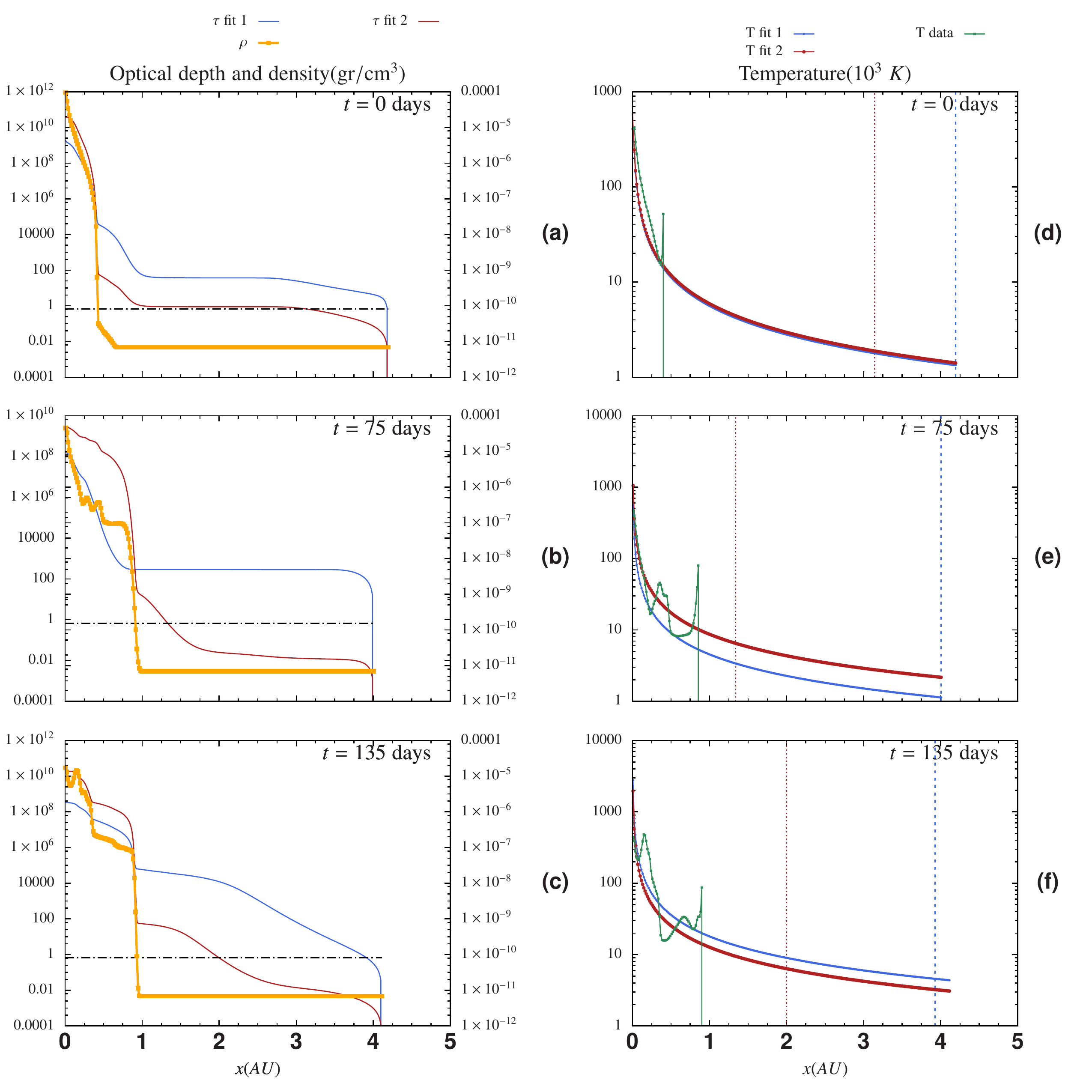}
     \caption{Left column: the density profile (yellow line, left vertical scale) at three times in the simulation along a ray in the $+x$ direction through the centre of the domain. Optical depths calculated using methods 1 (blue line, right vertical scale) and 2 (red line, right vertical scale) -- see text. The horizontal line marks the $\tau$=2/3 level. Right column: temperature as read from the simulation (green line - the upturn at larger radii is due to heating from the hot vacuum) and temperature fits using methods 1 (blue line) and 2 (red line) -- see text. The red and blue vertical lines mark the locations of the photosphere according to the two fits. The respective values of the effective temperature are listed in Table~\ref{tab:stratification}.}
  \label{fig:stratification}
\end{figure*}

A final attempt to resolve the problem of the determination  of the effective temperature was made by calculating a stratified, one dimensional atmosphere. This atmosphere could be effectively overlaid on the gas distribution and normalised at a location inside the photosphere, where we have confidence that the values of temperature and density are not affected by the vacuum temperature. By carrying out this exercise, however, we see that the choices are arbitrary and that the eventual value of the effective temperature has a severe uncertainty.

This  method  assumes  that   the  density  in the outer parts of the star follows  a  stratification structure with the following decaying power law:
\begin{align}
  \rho(r)=\rho_0\left ( \frac{r_{\mathrm{ph}}}{r} \right )^{2}, \label{eq:1a}
\end{align}
where  $\rho_0=\rho(r_{\rm ph})$ and  $r_{\rm ph}$ is a grid  point at the photosphere. Assuming that the exterior of the star behaves like an isentropic ideal gas with  
\begin{align}
  P \rho^{-\gamma}=&K, \label{eq:2a}\\ 
  P \rho^{-1}=&\frac{\overline R}{M}T, \label{eq:3a}
\end{align}
where $P$, $\rho$ and $T$ are the pressure, density and temperature of
the fluid respectively;  $M$ and ${\overline R}$ are the molar  mass and universal gas
constant,  respectively and  $K$ is  a  constant  to  be defined. Once again the
adiabatic index is $\gamma=5/3$.

From Eq.~\ref{eq:1a}, \ref{eq:2a} and \ref{eq:3a}  we can derive an  expression for the
temperature within the outer stellar gaseous layers:
\begin{align}
  T=&T_0 \left ( \frac{r_{\mathrm{ph}}}{r} \right )^{4/3}, \label{eq:4a}\\ 
  T_0=&\frac{M K \rho_0^{2/3}}{\overline R}. \label{eq:5a}
\end{align}
In Fig.~\ref{fig:stratification}, left column, we plot the density profile from the simulations alongside the optical depth that is calculated using that density, the opacity tables and values of the temperature that are calculated using the two following methods. In the first method, for every ray and every time, we used the simulation data and selected a cell just inside the photosphere, where we estimated that the value of the temperature was not affected by the hot vacuum. For this cell, we then read the temperature and position values and used these values as $\{T_0,r_{\mathrm{ph}}\}$ in Eq.~\ref{eq:4a}, thereby calculating values of $T$ for every value of $r$ outside the location of $r_{\rm ph}$.  Alternatively, a second method was to use a set of values $\{T^*_0,r^*_{\mathrm{ph}},\rho^*_0\}$ from the simulation at time zero to  calculate  $K$  in Eq.~\ref{eq:5a}  and  then use that value of $K$ to calculate the temperature at other points and other times, anchoring Eq.~\ref{eq:4a} at a point inside the photosphere at which we know the density, $\rho_0$, and the coordinates, $r_{\rm ph}$. 

The first method (red line in Fig.~\ref{fig:stratification}, right column) assumes that the temperature of the simulation is accurate
inside the star. However, given the high temperature vacuum, we
selected only cells with a negative gradient of the temperature profile (in Fig.~\ref{fig:stratification} these points are at 0.37, 0.54 and 0.79 AU). 
On the other hand, the second method uses the density and the
value of $K$, which is calculated with initial data only. Therefore, in
this case we assume that the density is accurate inside the star and
the temperature of the atmosphere follows Eq.~\eqref{eq:1a}.

Finally, the values of density needed to calculate the optical depth can be taken from the data directly, or, more self consistently with the stratification method, using Eq.~\ref{eq:1a}. We tested both cases. The results are similar. Therefore, in Fig.~\ref{fig:stratification} we present results obtained using the numerical values of the density only.  

As can be seen, the optical depth reaches a value of 2/3 at two different locations for the two methods. At those locations, the values of the temperature can be read from the right hand side column of Fig.~\ref{fig:stratification} and they are listed in Table~\ref{tab:stratification}. As is clear, these values vary greatly depending on the method followed, demonstrating that the values are arbitrary.

\begin{table}[h]
  \begin{center}
\caption{Effective temperature and photosphere location.  }\label{tab:stratification}
\begin{tabular}{ccccc}
    \hline  \hline
       Time &  \multicolumn{2}{c}{fit 1}  & \multicolumn{2}{c}{fit 2}   \\
       (days) &  $T_{\mathrm{eff}}\; (K)$ & $r_{\tau=2/3}$ (AU) & $T_{\mathrm{eff}}\; (K)$ & $r_{\tau=2/3}$ (AU)\\
       \hline
       0 & 1337 & 4.2 & 1884 & 3.1 \\
       75 & 1132 & 4.0 & 6480 & 1.3 \\
       135 & 4579 & 3.9 & 6351 & 2.0 \\
    \hline
\end{tabular}
\end{center}
\end{table}

\subsection{A provisional solution to determine the effective temperature}

During our hydrodynamic simulations, the fluid starts in the optically
thick regime, but as the gas  expands it may become optically thin,
at which point the photosphere  and  its  temperature   can  be  more  easily  determined
(although  in  the part of the simulation  presented  in  this  paper, the  fluid
remains  optically thick).   Normally,  the simulation  begins with  a
single star model with known effective temperature and luminosity from
the  original 1D  model.  As the companion starts its  infall through the primary stars' gaseous
layers the optically  thick photosphere expands. As long as the gas distribution remains fully optically thick and the temperature of the photosphere
is ill-defined (as explained in Sec.~\ref{ssec:resolutionproblem}) we make the following approximation: the first  opaque grid
point  is reassigned  to be  the $\tau=2/3$ surface  point with  temperature
$T=T_\eff$, unless the  temperature at the centre of the  cell is lower
than the effective temperature of the  initial model, in which case we
use the actual value:

\begin{equation}
T = \left \{
\begin{array}{lll}
T_{\rm eff} & \mathrm{if} & T \geq T_{\rm eff} \\
T & \mathrm{if} &T <  T_{\rm eff}\\
\end{array}
\right.,  \label{eq:15}
\end{equation}
\begin{figure*}
  \centering
  \includegraphics[width=100mm]{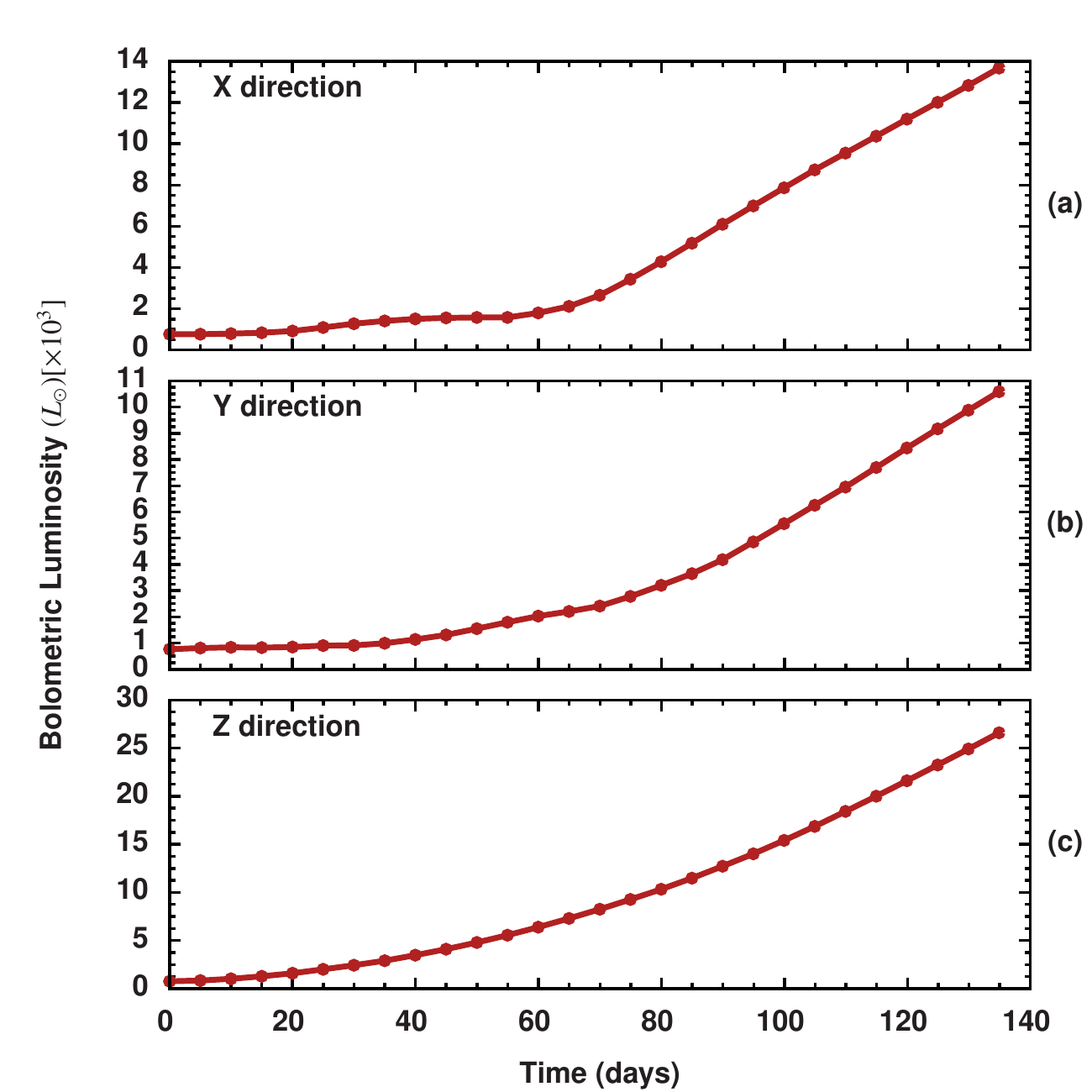}
  \caption{The  bolometric  luminosity as  function  of  time for the highest
    \Enzo\    resolution available  ($N=512$)
    towards each of the three directions}\label{fig:5}
\end{figure*}

\begin{figure}
  \centering
\includegraphics[width=85mm,clip=true,trim=0mm 0mm 0mm 0mm]{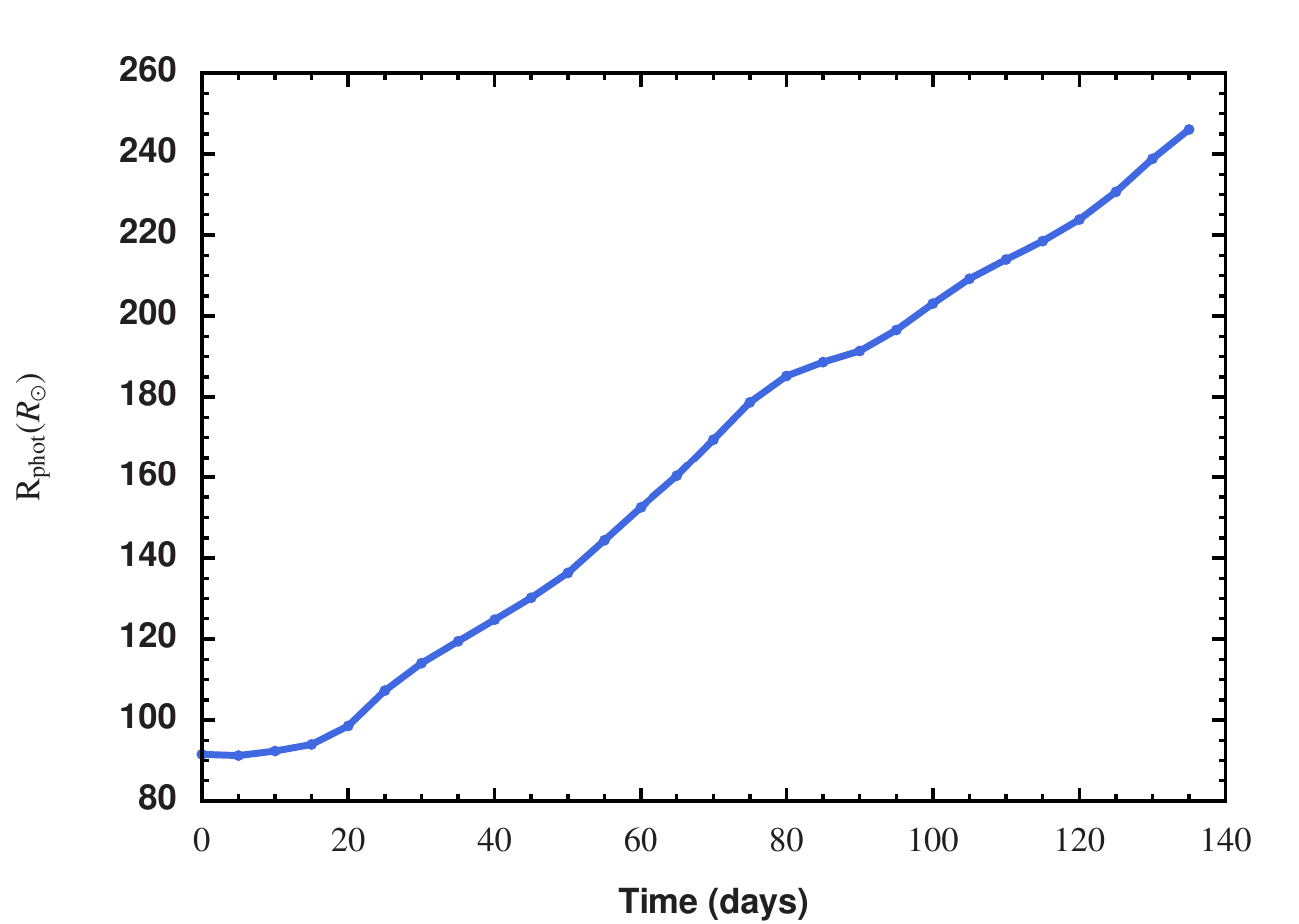}
     \caption{The radius of a sphere that has the same volume as the photospheric surface.}
  \label{fig:radius}
\end{figure}

This implicitly assumes that at $t>0$ the temperature of the expansing photosphere {\it decreases}, something that, as we will see in Section~\ref{sec:observations} is not always the case.

We also ensure that the combination of temperatures used does not lead
to a  total luminosity  smaller than  the initial  stellar luminosity,
since the  stellar luminosity is provided  by a thin shell  resting on
the core of  the primary and the CE interactions  we are modelling are
not thought to alter the nuclear burning rate on short timescales.

For the short time over which the photosphere remains optically thick,
using a constant value of the effective temperature is likely correct to better than a factor of two, since the temperature is regulated by  the opacity
and there is no time for radiative cooling. However, this is still a problematic assumption and the single largest challenge in determining the light from this type of simulation. We discuss this further
in Section~\ref{sec:discussion}.

\section{Results: Towards calculating the light curve of a common envelope simulation} \label{sec:results}

\begin{figure*}
  \centering
  \includegraphics[width=120mm]{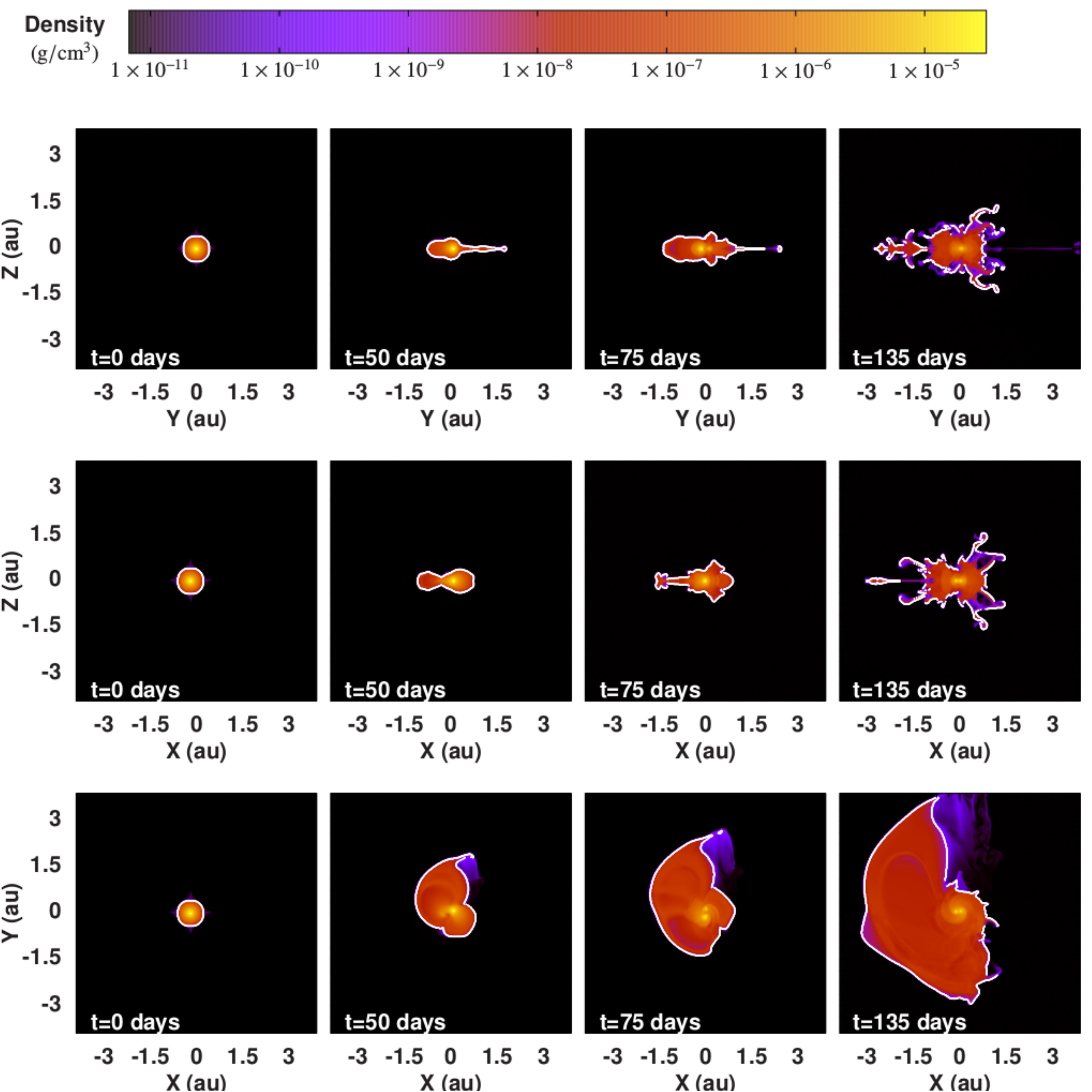}
  \caption{Density slices located at $X=0$, $Y=0$, and $Z=0$ (from top to bottom), taken at  times
    $t = 0$~(column\,1),  50~(column\,2), 75~(column\,3), and 135~(column\,4)~days. The white line
    represents  the  photosphere as located by an observer on the same plane as the density slice.}\label{fig:6}
\end{figure*}

In this section we show the  light curve for well studied CE evolution
simulations: Enzo2 from P12, which we have carried out with a  computational domain four times as large and the same resolution, as explained in Sec.~2. In  Fig.~\ref{fig:5}  we show  the  bolometric  light  as seen  by  an observer located  along three  orthogonal directions, parallel  to the $x$, $y$ and $z$ axes, while in Fig.~\ref{fig:radius} we show the volume-equivalent radius of the photosphere. We emphasise that the values of the temperature are almost always those of the initial $T_{\rm eff}$ of the star (3200~K, see Table 1 in Appendix) because the values of the photosphere almost never drop below this value during the early, optically-thick photospheric expansion. This effectively means that we are assuming a constant temperature photosphere. In Fig.~\ref{fig:6} we show density slices both on  the  orbital  and  perpendicular  planes.  During  the  entire  CE evolution, the model remains effectively optically thick. In addition, as  soon as  stellar material  leaves  the domain  the photosphere  is
effectively lost. Despite our  calculation with a larger computational domain, this happens  at approximately 135 days, which is short of the $\sim$200 days taken  by the fast in-fall phase and  even shorter than the $\sim$1000 days of the entire simulation run of P12.

Throughout this  entire CE simulation, the  photosphere coincides with
the density of the \Celmo\ density floor. However, lowering this floor
further does not change the light  output because of the steep density
gradient at  the photosphere. A  very small difference might  be found
toward the end of the simulation time. As can be seen in Fig.~\ref{fig:6},
at 75  and 135 days, some  material extends past the  photosphere. This
gas has a density intermediate  between the \Enzo\ and \Celmo\ density
floors   and  could   be  optically   thick,  thereby   extending  the
photospheric area slightly.  However, in our simulation  this very low
density stellar gas has a temperature that is greatly increased by the
low density,  high temperature vacuum  medium and therefore  cannot be
studied.

\begin{figure*}
  \centering
  \includegraphics[width=100mm]{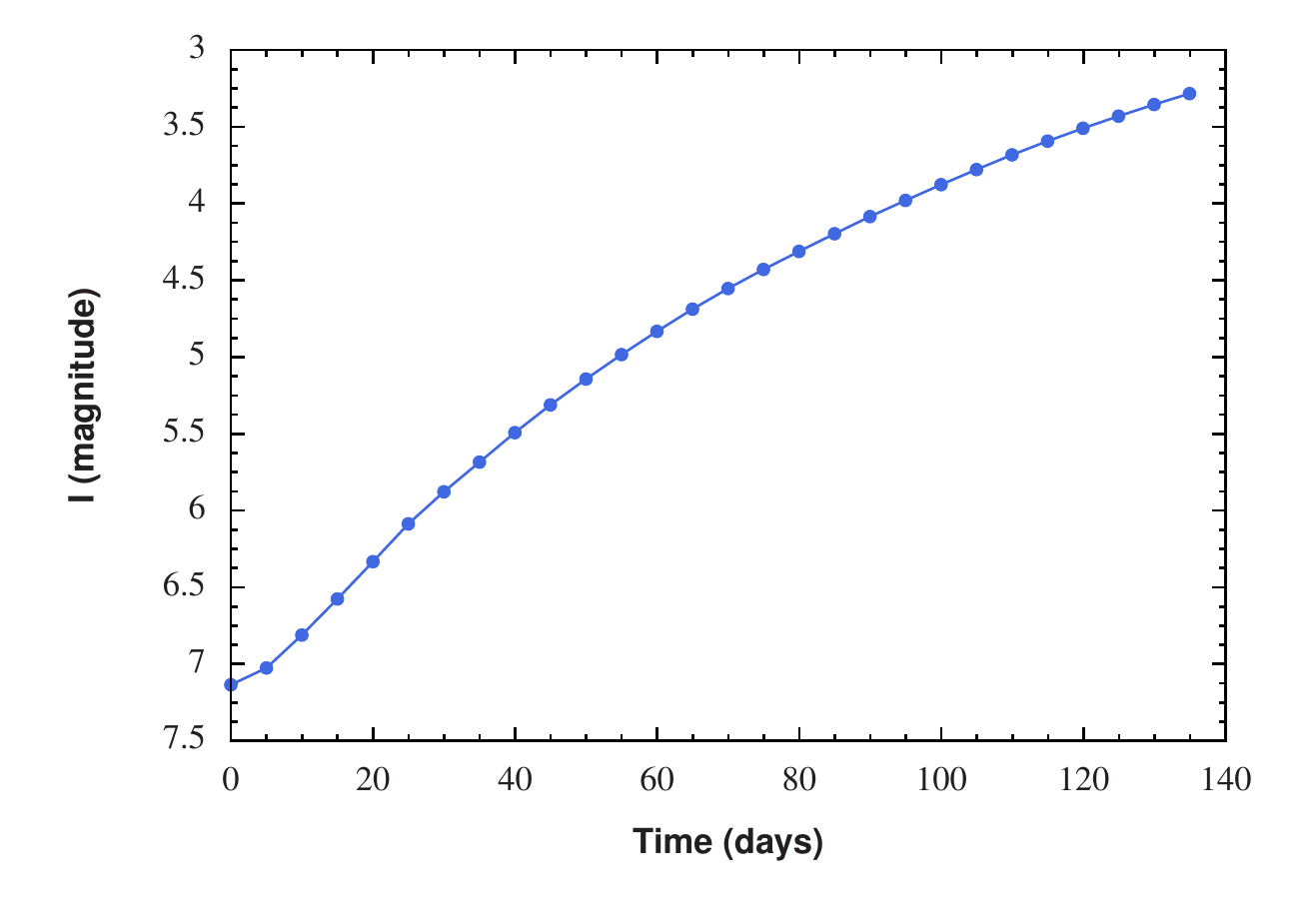}
  \caption{$I$-band magnitude of the system observed along the $z$ direction (looking perpendicularly to the orbital plane), as if the system were observed from 
    1 kpc with no reddening.}\label{fig:7}
\end{figure*}

In Fig.~\ref{fig:7} we show the $I$ band luminosity of one of the two
perpendicular  views, setting  the object  at 1~kpc  and including  no
reddening. The  calibration values to  derive the magnitudes  from the
luminosities  are those  found  in  Appendix~\ref{ssec:filter}. The  $V-I$
colour of the  initial model is 1.98 and would  mostly not change over
time since  the photospheric temperature is  effectively constant. The
initial rise of  the $I$ band luminosity is almost  3 magnitudes in 135
days.

The  total radiated  outburst energies  between the  beginning of  the
simulations and 135 days are $3.7 \times 10^{43}$, $3.5 \times 10^{43}$
and  $9.6 \times  10^{43}$~erg for  the $x$,  $y$ and  $z$ directions,
respectively. We compare  these energies with the total  energy in the
AGB star at  the start of the simulation (namely  its thermal, kinetic
and potential  energies) which is  $\sim -2 \times  10^{46}$~erg. This
validates  the adiabatic  approach  for the  short  timescale we  have
simulated here.

\section{Guidance from Observations} 
\label{sec:observations}

\subsection{Comparisons with transients}
At least three transients have been credibly identified as CE interactions, primarily because their progenitors were observed: V1309~Sco \citep{Tylenda2011}, M101-OT \citep{Blagorodnova2016}, and M31-2015~LRN \citep{MacLeod2016}. Due to  similar light and spectral characteristics after the outburst, other transients, such as V~838~Mon \citep{Bond2003} or NGC4490-OT \citep{Smith2016b} have been suggested as having a similar origin. Here we carry out a comparative discussion of those aspects of the observations that today or in the near future will be the most useful to constrain CE simulations. We also highlight those aspects of the simulations that will be best constrained by observations. We concentrate on M31-2015LRN for which \citet{MacLeod2016} have extracted system parameters from their observations.

V~1309~Sco is in a way the system that is the closest to our simulations, in light of its low mass, a  $\sim 1.5$~\msun\ subgiant interacting and merging with a  $\sim 0.15$~\msun\ companion \citep{Tylenda2011}. M31-2015LRN (and likely V~838~Mon) is possibly next, in terms of system's mass: a $3-5.5$~\msun\ primary interacted with a $0.1-0.6$~\msun\ companion and thought to have undergone a CE merger event. M101-OT was instead thought to come from the interaction between an 18-\msun\ primary and a 1-\msun\ companion, with NGC4490-OT being even more massive, though an actual value of the mass could not be derived \citep{Smith2016b}. The peak absolute luminosities of these outbursts are listed in Table~\ref{tab:observations}, alongside their lightcurve behaviour. 

\begin{table*}
\centering
\begin{tabular}{lcccll}
\hline
Name &  $M_{\rm Band}$(peak) &Band & $M_1$&$M_2$&  Plateau? \\
          &(mag)             &&(\msun)&(\msun)&\\
\hline
Simulation$^1$        & $<-$4.8::, $<-6.8$ &$V,I$& 0.9   &0.6      &--\\
V~1309~Sco$^2$    & $-6.9$,$-7.9$        &$V,I$& 1.5   & 0.15    &n?\\ 
V~838~Mon$^3$     & $-9.6$     &$V$    & 5-10 & --         &n\\
M31-2015LRN$^4$ & $-9.5$     &$V$    & 3-5.5 & 0.1-0.6  &y\\
M101-OT$^5$         & $-12.6^a$&$r$    & 18&1 &y\\
NGC~4490-OT$^6$& $-14$       &unfilt.& 20-30:&--          &?\\
\hline
\multicolumn{6}{l}{$^1$P12 and this work; ``::" means very uncertain value.$^2$\citealt{Tylenda2011};}\\
\multicolumn{6}{l}{$^3$\citealt{Bond2003}; $^4$\citealt{MacLeod2016}; $^5$\citealt{Blagorodnova2016};  }\\
\multicolumn{6}{l}{$^6$\citealt{Smith2016b}. $^a$Could have been brighter, peak not observed.}\\
\end{tabular}
\caption{Some characteristics of observed transients interpreted as common envelope interactions/mergers.}
\label{tab:observations}
\end{table*}

From observations of transients, quantities such as the evolution of the photospheric radius, temperature and luminosity, as well as ejected masses, velocities and timescales of the various phases  can be determined, subject to some uncertainties such as on distance and reddening. \citet{MacLeod2016} used photometry of the M31-2015LRN outburst to deduce that the photospheric radius increased between 200 and 400~\rsun\ before peak brightness and then to 2000~\rsun\ in the next 30 days. The photospheric expansion velocity was measured to be 360~km~s$^{-1}$ from spectroscopy. They also determined that the photospheric temperature increased between $\sim 5000$~K and $\sim 7000$~K during the rise to peak (or $\sim 6500$~K to $\sim 11500$~K for the highest possible reddening value), followed by a steady decrease to $\sim 3000$~K in the next 50 days. Overall, the bolometric luminosity increased between $10^{38}$ and $2\times 10^{39}$~erg~s$^{-1}$ during the rise ($2 \times 10^{38}$ and $7\times 10^{39}$~erg~s$^{-1}$ for the highest reddening value) and declined to $\sim 5 \times 10^{38}$~erg~s$^{-1}$ in the next 50 days. 

From our simulations the most reliable quantity is the photospheric radius evolution over 135 days: the volume-equivalent radius (Fig.~\ref{fig:radius}) increased between 85 and 250~\rsun, recalling that this simulated radius may not have reached its maximum extension. Subject to the caveat of the uncertain temperature (Section~\ref{sec:lumin-calc}) the simulated bolometric luminosity goes from 648 to 14\,000~\lsun\ ($x$-direction) in 135~days ($10^{36}$ to $3 \times 10^{37}$~erg~s$^{-1}$). Our progenitor's absolute magnitudes are $M_{\rm I,prog}=-2.9$ and $M_{\rm V,prog=-0.9}$ (using the average $V-I$ value calculated above), while at 135~days we measure $M_{\rm I,135d}=-6.8$ and using the same color correction we obtain a value of $M_{\rm V,135d} = -4.8$. 

Lacking at present the ability for a direct comparison, we can however still place the simulated values of $M_V$ and $M_I$ for the progenitor and for the expanded star on the mass vs. absolute magnitude plot of \citet{Kochanek2014}, who showed that the more massive systems have brighter outbursts. As pointed out by \citet{Smith2016b} this, and the fact that more massive progenitors have longer outbursts, could be explained by the fact that more massive progenitors have more kinetic energy and angular momentum and longer radiation diffusion times. Using the fits of \citet{Kochanek2014}, we would expect a 0.9~\msun\ progenitor to have $M_{\rm V,prog}=4.7$ and $M_{\rm I,prog}=3.5$. Our progenitor is brighter and redder than predicted by the fit of \citet{Kochanek2014}, likely because our star is evolved, while the fitted data are for unevolved stars. In fact OGLE-2002-BLG-360, also plotted by them, but not fitted, is a more evolved star  and is indeed brighter. Their fits would predict that a 0.9~\msun\ unevolved star would have an outburst with peak brightness $M_{\rm I, peak} = -5.3$ and $M_{\rm V, peak} = -3.4$. Comparing their predicted $I$ band with ours (Table~1) shows that our magnitude is at least 1.5 mag brighter, though the $V-I$ colours are similar. This is at this stage acceptable in view of the many uncertainties.

\subsection{Constraints from Mira giants}
The temperature of the photosphere and the luminosities are not well constrained; as explained,  the temperature is effectively kept constant at the value of the progenitor's effective temperature.  Mira variables are AGB giants with characteristics similar to the stars we have considered in our simulation. They expand due to pulsations on timescales similar to the expansion timescales considered here and in so doing their radius changes similarly to the radial expansion considered here. In models of $o$~Ceti \citep{Ireland2008,Ireland2011}, the effective temperature of the photosphere changes between 3800~K and 2200~K during half a pulsation cycle of 330 days (i.e., 165 days). During this time the radius expands by a factor of 2.3.  This is similar to what was found for other Mira stars. Our calculation  over 135 days sees an approximate radial expansion over a similar factor. Such a decrease in temperature would give a reduction in luminosity by a factor of $\sim$10 compared to the values we have estimated.  

On the other hand, the expanding photosphere may {\it not} initially cool. In the case of M31-2015LRN, the expanding photosphere was initially {\it heated} by shocks, instead of cooling adiabatically by expansion, and only later cooled. Therefore, observations caution us that assuming that the photosphere initially cools by expansion may be misguided.

\subsection{Recombination energy as an agent in the the common envelope ejection}
Another related issue of fundamental importance is that of recombination energy released upon recombination of hydrogen and helium. The release of  recombination energy as light may explain the plateau in certain type of Type II supernovae \citep{Ivanova2013b}. \citet{MacLeod2016} argue that the plateau in the light curve of the transient M31-2015LRN after the maximum is due to such energy release. 

However, \citet{Ivanova2015} and \citet{Nandez2016} argued that during the common envelope expansion, recombination energy is released at such high physical depth that the optical depth should also be large, making the energy released there entirely available to generate pressure that results in the expulsion of the CE. At such depth, they argued, even the dramatic decrease in opacity of recombined gas may not be sufficient to liberate the energy as light on short timescales and is therefore available to do work. This is a very important point that needs a resolution: if recombination energy does not escape, then it must be included in CE simulations, but if part or all of it escapes, then CE simulations that include recombination energy and that are run in the {\it adiabatic} approximation, will overestimate the ejected mass, ejecta's speeds, and produce unphysical in-spiral behaviours \citep[see][for why expanding, recombining giants do not blow themselves apart]{Harpaz1998}. Observations such as those listed here, and particularly the presence or absence of a plateau in the lightcurve, may point to an observational constraint on how recombination energy is transformed in the star.

\subsection{What happens just before the common envelope in-spiral?}
Another aspect of the interaction where observations will provide us with a quantitative constraint concerns what precedes the fast in-spiral. \citet{MacLeod2016} suggested that there can be two pre-in-spiral scenarios with distinct observational characteristics. The first, which they apply to M31-2015LRN, is a fast (days) pre-in-spiral phase where a secularly stable orbit is destabilised by the Darwin instability \citep{Darwin1879}. This leads to Roche lobe overflow and the CE in-spiral in quick succession. This phase is characterised by an early ejection of a low mass, but optically thick shell that is observed as an expanding photosphere. At the same time the photospheric temperature increases as this gas is shock-heated by the early in-spiral. This shell is ejected with speeds above escape velocity. Right after this ejection the full photosphere expands and cools, driven by orbital energy deposited during the in-spiral. The second scenario, is a slower one: after Roche lobe overflow the mass transfer remains stable and leads to an outflow from the second Lagrangian point at lower ejection speed (25\% of escape speed). The expanding gas creates a wall of material into which the subsequent expansion phase, driven by the in-spiral, will collide. The difference between these two scenarios is key to understanding when a CE is avoided.

The simulation presented in P12 and this paper cannot help chose between the two scenarios because the companion is placed on the surface of the giant at the start of the simulations and the primary therefore already well exceeds its own Roche lobe radius. However, one of the SPH simulations presented by \citet{Iaconi2017} may afford a better comparison. That simulations is identical to that of P12 analysed here, but it started with a wider orbital separation, with the primary at Roche lobe contact. The stable mass transfer phase, preceding the fast CE in-spiral, lasts a decade, but this is likely a lower limit \citep[][and Reichardt et al., in preparation]{Reichardt2015}. 

During this Roche lobe overflow phase, mass is ejected from the second Lagrangian point (L2, on the side of the companion) with speeds of 100-150~km~s$^{-1}$, which is above the local escape speed of $\sim$60~km~s$^{-1}$, while at the third Lagrangian point (L3, on the side of the primary), gas is being ejected with speeds of $\sim$40~km~s$^{-1}$, which is similar to the local escape speed of $\sim$50~km~s$^{-1}$. The expansion of the photosphere measured here is between 85 and 250~\rsun\ and lasts 135 days, implying an expansion speed of 9~km~s$^{-1}$ lower than the escape speed and lower than the speed of the ejecta seen emerging from L2 and L3. Once again is it difficult to make quantitative comparisons at this time, but as these values become refined, they will be those that allow us to discriminate what happens before the CE in-spiral and how this phase affects the CE proper and the post-CE parameters.

CE interactions and merger observations assume that the expanding primary was caught in a CE right after the main sequence as the star commenced its journey towards the red giant branch. However, statistically, the rate of interactions involving more evolved giants should be larger, because there are more companions at larger separations \citep{Duchene2013}. As more transients and their progenitors are observed, we will know whether the apparent overabundance of sub-giant branch mergers is due to their being particularly bright or simply to our current uncertainty on the nature of the progenitors. If the former, an explanation could be that the CEs we observe are those with a more bound envelope (more massive and compact) where the companion penetrates deeper and tends to merge more readily, releasing more energy.

\subsection{Dust formation during common envelope expansion}
Finally, observations tell us that dust will have to be considered in the calculation of the light, as it may deeply affect the lightcurve. As early as during the first few days 
of the expansion, dust could be produced, as seen in V~1309~Sco  \citep{Tylenda2011} that  brightened
over a period of approximately 5 years and then suddenly dimmed by about a
magnitude (in the $I$ band) just  before the outburst. \citet{Nicholls2013} obtained an IR spectrum approximately a year
after  the optical  outburst peak  and determined  that a  substantial
amount  of  dust  must  have   condensed  in  this  object,  though  a
determination of the dust mass was impossible without knowledge of
the  dust geometry.  This system  was known  to be  a close-to-edge-on
contact binary. We conjecture that the  dust forms in a disk along the
equator. This  is in line with  the large dust grain  size measured by
\citet{Nicholls2013}        that       necessitate        a       disk
environment. Additionally, \citet{Chesneau2014}  measured an elongated
dusty environment,  interpreted as a  disk, in V838~Mon. We could
therefore conjecture that while dust formation during the CE in-spiral
is possible,  indeed likely, it  may primarily influence the  light as
seen  along  the  orbital   plane,  leaving  perpendicular  viewpoints
relatively unobstructed.

\section{Summary and Future Work} \label{sec:discussion}

In  this article,  we  presented a  computational code  to
post-processe   the   luminosity    for   hydrodynamic   simulations.
\Celmo\  is  currently designed  to  compute  the light-curve  for  CE
simulations  performed  with the  \Enzo~  code  in unigrid  mode.  We
presented a  first attempt at  calculating the light-curve for  one of
the  CE  simulations  presented by  P12,  comprising  an
$\sim$80-\rsun, 0.88-\msun, RGB star with a 0.6-\msun\ companion.

The computation of the light from CE interactions is paramount if we 
are to use current and upcoming observations to constrain simulations. Our attempt
cannot at this time be considered satisfactory, because we have maintained the photospheric temperature constant over the 135 days of the simulation. With this effort we have, however,  
elucidated and quantified the main issues with the computation. This is a fundamental step before a solution can be determined. Below we summarise these shortcomings of the current attempt and discuss possible avenues towards a solution, some of which we will explore in future papers in this series.

The limitations of  our approach can be divided into  two groups. The
first group  includes limitations  inherited from the  3D hydrodynamic
computation  itself,   while  the  second  group   comprises  physical
limitations due to simplifications in the light calculation.

The main computational limitation are the impossibility to resolve the value of the photospheric temperature during the optically thick phase of the expansion, and even more importantly, the heating of the outer layers by the hot ``vacuum". These make it impossible to read a photospheric temperature value directly from the simulation. According  with the  values of  opacity and density  of our  models,  we would  be able  to  properly resolve  the
photospheric temperature, only with a cell size smaller than 0.004~\rsun\ (cf.  with
3.4~\rsun\ for our simulation reproducing $Enzo2$ of P12). The best resolution attained to date by the simulations of Ohlmann et al. (2016) is 0.01~\rsun\ at the centre of the domain. Hence, this is at  the edge of our capabilities even  with an adaptive mesh refinement (AMR) code, 
as it would require between 8 and 9 levels of AMR refinement over the 
large volume occupied by the photosphere. Even with higher resolution, however, the problem of the external artificial heating remains.

None of the techniques we have tried to reduce the uncertainty on the effective temperature  has proven satisfactory. Aside from a massive improvement in the resolution of the original calculation, which is not within immediate reach, we are exploring a way to interface \Enzo\ and {\sc rage} \citep{Gittings2008} in order to exploit the latter code radiation capabilities, while using information from the \Enzo\ computation. 

Another computational limitation is one  of the  largest issues  with 
hydrodynamic  computations of stars using grid codes is the finite size of the domain, which allowed us
to compute the light for only 135 days of the CE simulation of P12. 
This problem  will  be greatly  alleviated  by  using  the AMR  version  of
\Enzo\  \citep{Passy2014},  which will  allow  us  to  maintain
resolution with a much larger box. 

Our smooth particle hydrodynamics simulations \citep{Iaconi2017} could also be used. They do not have a hot vacuum and have no computational domain limits, but at present the resolution near the photosphere is even lower than for the grid simulations due to the low density in those regions and to the computational times that at present limit the resolution to approximately a million particles.

Once the  technical issues  listed above have been resolved,  we 
will  have to  contend with  physical ones.  During the  initial
infall the timescale is dynamical.  This is the reason why simulations
of this  phase can  avoid the  inclusion of much  of the  physics that
would  dominate  over  longer  timescales. Radiative  cooling  is  not
expected on such short timescales (as  also confirmed by the fact that
the total energy radiated over the  initial 75 days of the interaction
is  much  less than  the  initial  energy  of  the envelope)  and  the
thermodynamic properties of the gas  should be well represented by our
adiabatic  calculation. Yet,  later  on, we  can  expect cooling,  gas
recombination and  the formation of  molecules and of dust.  This will
alter  the  opacities  and  necessitate  the  treatment  of  radiation
transport. However,  we note that predicting
the initial lightcurve rise may not necessitate the treatment of these processes.


\section{Acknowledgments} 
\label{sec:acknowledgments}
It is  a pleasure to  thank Juhan  Frank and Mark Wardle for valuable  discussions and
comments on  the manuscript. Peter Wood is thanked for discussions regarding
the properties of Mira stars. An anonymous referee is thanked for extensive comments that helped significantly the maturity of this manuscript. Thomas Reichardt is thanked for sharing some details of his upcoming publications regarding ejection speed of mass from CE interactions. This work  was supported  by the
Australian  Research  Council   Future  Fellowship  (FT120100452)  and
Discovery Project (DP12013337) programmes.

This research was undertaken on the NCI National Facility in Canberra,
Australia,  which   is  supported   by  the   Australian  Commonwealth
Government. 
This work  was performed on the  swinSTAR supercomputer at
Swinburne University of Technology.
Computations described in this work were performed 
using the Enzo code\footnote{http://enzo-project.org},  which is  the product  of a  collaborative
effort of scientists at many universities and national laboratories.

JCP thanks  the Alexander von Humboldt-Stiftung and Norbert Langer
for their support.


\appendix

\section{The formulation} 
\label{sec:formulation}

\begin{figure}
  \centering
  \includegraphics[width=84mm]{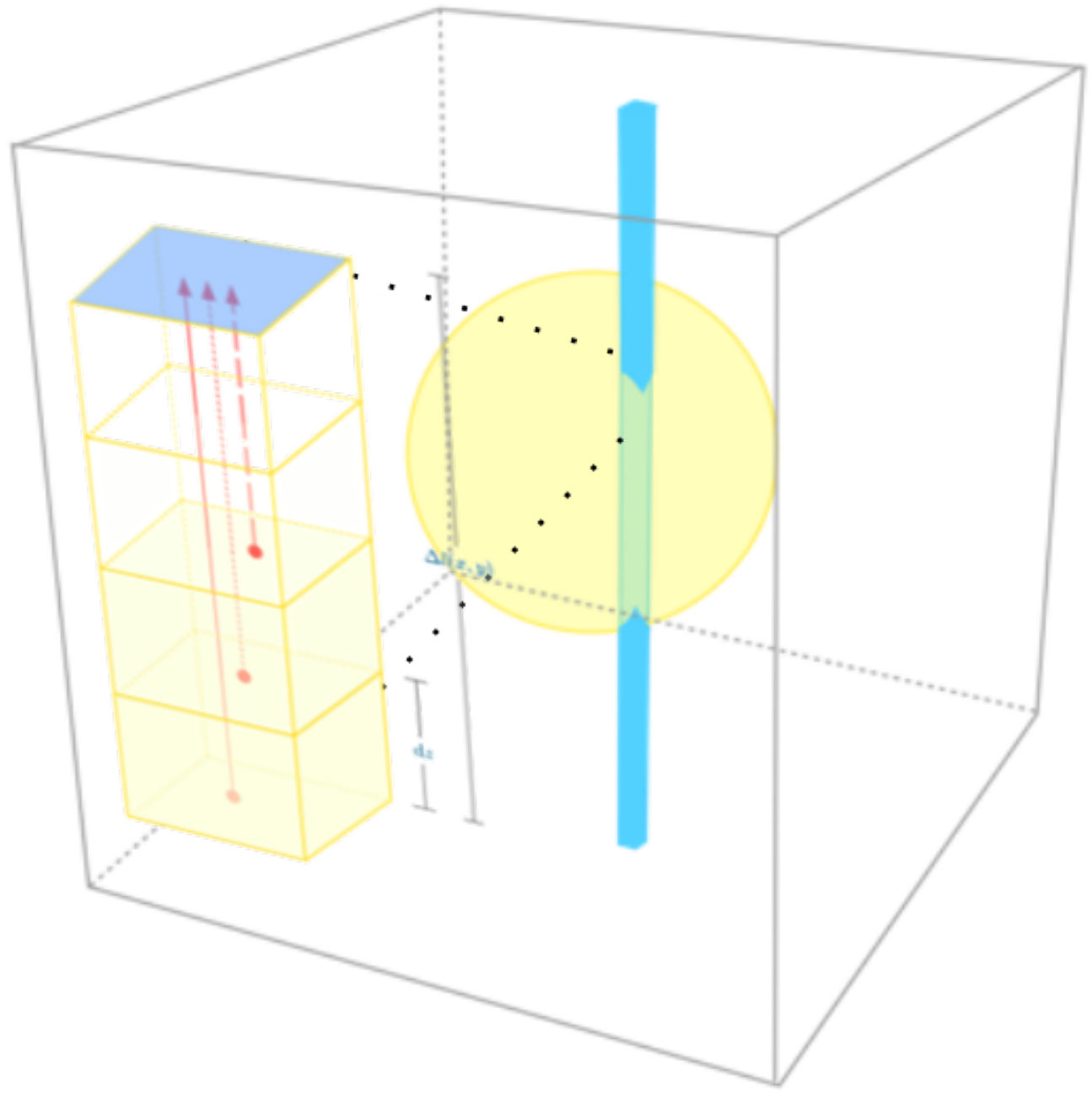}
  \caption{ Numerical domain partition. The numerical domain is divided
    in columns along the line of sight. The light is computed for each
    volume  element   and  the  total  contribution   of  each  column
    integrated for  each face of  the numerical domain  (diagram shows
    the case for $+Z$ propagation) }\label{fig:1}
\end{figure}

Each numerical volume element, or cell, in our simulation has a temperature $T(x,y,z)$ and density
$\rho(x,y,z)$.   If  we  observe  the computational  domain  from  the
positive direction  of the $\hat{z}$ axis, with  the coordinate system
origin   in  the   centre  of   the  numerical   domain   (see  Fig.
\ref{fig:1}), we  can then divide  the cube into columns  of area
$a=\Delta x  \Delta y$ and infinitesimal  depth of width  $\rd z$. The
surface brightness of each slice is:

\begin{equation}
\rd B(\nu,\vec{r};T) = \frac{N(\nu,z;T)h \nu }{a \omega}\frac{\rd z}{\Delta l(x,y)}  \quad , \label{eq:1}
\end{equation}

\noindent                  in                 units                 of
erg~s$^{-1}$~cm$^{-2}$~Hz$^{-1}$~sr$^{-1}$,  where $\omega$ 
reppresents the solid angle subtended by the area $a$ at the observer, 
and $N$ denotes
the number of  photons of frequency $\nu$ emitted  from the surface of
area $a$.  The energy of each photon is $\h \nu$ and has an associated
temperature  $T$.   The fraction  $\rd  z/\Delta  l(x,y)$ denotes  the
proportion $\rd  z$ of the total  length of the column
$\Delta  l(x,y)$ and  is equal  to  $l(x,y)_{\max}-l(x,y)_{\min}$ (see
Fig. \ref{fig:1}). The specific intensity:

\begin{equation}
I(\nu;T)= \frac{N'(\nu;T) \h \nu}{A \Omega}, \qquad \label{eq:2}
\end{equation}
 
\noindent  (with  the same  units  as  Equation~\ref{eq:1}), is  the
number of  photons $N'$ with energy  $h \nu$ detected at a surface of
area $A$ arriving  from solid angle $\Omega$.  The number of photons
arriving at the detection surface  is related to the number of photons
emitted by:

\begin{align}
N'(\nu)=&\int_{l_{\min}}^{l_{\max}} N(\nu,z)\e^{-\tau}\rd z , \label{eq:3}  
\end{align}
\noindent where:
\begin{align}
\tau:=& \int_z^{l_{\max}}\kappa(x,y,\xi)\rho(x,y,\xi)\rd \xi , \label{eq:4}
\end{align}

\noindent where $\kappa$ and $\rho$ are the oapcity and density of the
medium, respectively,  and $\tau$ is the optical depth.   The exponential
factor takes  into account the  dispersion of photons between  $z$ and
the surface of  the volume $l_{\max}$. The integration  in $z$ gives us
the total contribution of the column.

Assuming blackbody  radiation, the surface brightness of  one slice in
our domain is:

\begin{align}
B^*(\nu;T)_{ijk} & = 2\frac{h \nu^3}{c^2}\frac{1}{\e^{h \nu/k T_{ijk}}-1}\label{eq:5}
\end{align}
where $i,j,k$ are the $x,y,z$ discrete indices of the numerical domain. 
The specific intensity in the respective column is:

\begin{align}
I(\nu;T)_{ij} = \sum_k  B^*(\nu;T)_{ijk}\; \e^{-\int_{z_{ijk}}^{l_{\max}}\kappa(\xi)\rho(\xi)\rd \xi} \label{eq:6}
\end{align}

The radiation flux crossing the detection surface is given by:

\begin{align}
S(\nu;T)_{ij} &=  \int_0^{2\pi} \int_0^{\pi/2}I(\nu;T)_{ij} \cos \theta \sin \theta \rd \theta \rd \phi \label{eq:7} \\
&= \pi I(\nu;T)_{ij}, \qquad   \label{eq:8}
\end{align}

\noindent   where   we   have assumed   that   the   observation   distance
$z_{\mathrm{obs}}>>\max(\Delta l(x,y))$ and isotropic emission of each
column. The flux density $\mathcal{F}$ is given by:

\begin{equation}
\mathcal{F}_{ij}= \int f(\nu) S(\nu)_{ij} \rd \nu, \qquad , \label{eq:9}
\end{equation}

\noindent in  units of erg~s$^{-1}$~cm$^{-2}$ where $f(\nu)$
represents a  filter (see  Sec. \ref{ssec:filter}). The  luminosity is
therefore:

\begin{equation}
L= \sum_i \sum_j \mathcal{F}_{ij} \Delta x \Delta y. \qquad \label{eq:10}
\end{equation}

\subsection{Convolution with filter band passes} 
\label{ssec:filter}

Light  curve observations  are performed  in specific  spectral bands.

Assuming homogeneous emission in  all frequencies, each volume element
should radiate  in the full  spectrum. Therefore, we can  convolve the
brightness of each slice with  the filter function. The  energy flux
density is:

\begin{align}
\mathcal{F}_{\mathrm{filt}}(T) &= 2\pi \frac{k^4 T^4}{c^3 h^3} \int_0^\infty \frac{f(\chi) \chi^3}{\e^{\chi}-1} \rd \chi \label{eq:9}\\
&= \frac{15 \sigma}{\pi^5} T^4 \int_0^\infty \frac{f(\chi) \chi^3}{\e^{\chi}-1} \rd \chi, \label{eq:10}
\end{align}
where $\chi = h  \nu / k T$, and $h$, $k$, $c$ are the Planck constant, the Boltzmann constant and the speed of light, respectively.

The bolometric flux density is recovered if $f(\chi)=1$, or:

\begin{align}
\mathcal{F}_{\mathrm{bol}}(T) &= \sigma T^4. \label{eq:11}
\end{align}
For  a   star  with  effective   temperature  $T_{\mathrm{eff}}$,  the
 bolometric luminosity is  $L_{\rm bol} = 4\pi
R_{\mathrm{star}}^2 \sigma T_{\mathrm{eff}}^4$.

\noindent The effect of the filter is encoded in the factor
\begin{align}
f_{\mathrm{filt}}(T)&= \frac{15 }{\pi^5} \int_0^\infty \frac{f(\chi;T) \chi^3}{\e^{\chi}-1} \rd \chi.\label{eq:12}
\end{align}
Therefore, the energy flux density $\mathcal{F}_{\mathrm{filt}}(T)$ takes the form
\begin{align}
\mathcal{F}_{\mathrm{filt}}(T) &= \mathcal{F}_{\mathrm{bol}}(T) f_{\mathrm{filt}}(T). \label{eq:13}
\end{align}
We  integrate numerically Equation~\eqref{eq:12} using
the SciPy\footnote{www.scipy.org} routine of Simpson's rule.  
To compute  the magnitude, we  use the bolometric  magnitude, $V$-band
magnitude   and   $V-I$    colour   of   the   Sun,   $M_{\sun, {\rm bol}}=4.75$,
$V_{\sun}=-26.74$\footnote{nssdc.gsfc.nasa.gov/planetary/factsheet/sunfact.html}
and              $(V-I)_{\sun}=0.701$             \citep{Ramirez2012},
respectively. Additionally,  we use our  filters to compute  the solar
luminosity  in the  $I$ and  $V$ bands  using a  blackbody  curve with
$T_{\eff,\sun} = 5778$~K.

\subsection{Numerical implementation} \label{ssec:numericalimplementation}

\Celmo~  is  written  as  a  set   of  python  classes.   We  use  the
\textsc{SciPy}   library  for   integration   and  interpolation   and
\textsc{NumPy} for general mathematical operations. The purpose of the
classes  in  \Celmo~  are  the   interpolation  of  the  opacity,  the
convolution  of the  luminosity  with filter  functions  and the  main
calculation   of   the   luminosity    described   in   the   previous
section. Additionally, we use a  set of Python scripts to post-process
the data. \Celmo~ uses \YT~ \citep{TurSmiOis11} interface to read data
from \Enzo.

We  use the  \textsc{vtk} \citep{VTK}  data format  for 3D  output and
ASCII format for  1D and some of the 2D  quantities.  \Celmo~ requires
the density  field and either  the temperature or the  internal energy
(see  \S~\ref{sec:temperature}).   Additionally,  it is  necessary  to
specify the  effective temperature of the  initial model $T_{\eff}^0$.
The output fields are: the opacity  $\kappa$, the optical depth in six
directions  ($\pm x$,$\pm  y$,$\pm  z$), the  specific intensity,  the
energy  flux  density  in  each  direction,  the  luminosity  in  each
direction and the volume inside the $\tau=2/3$ surface.

We use  \textsc{MPI4py} \citep{DalPazDEl08} to take  advantage of multi-core
processors.  It  is straightforward  to  calculate  the luminosity  in a
distributed  computation  scheme  since  the necessary  quantities  are
located independently in the fluid columns.

\appendix
\section{Code verification}
\label{sec:verification}

In  this appendix  we  use two  simple models  which  are suitable  for
comparison with analytic expressions. The  easiest case is for a fluid
with constant opacity and density.

\subsubsection{Test I. Rectangular cuboid under optically thin conditions} \label{sec:rectangular-cuboid}

\begin{figure}
  \centering
  \includegraphics[width=84mm]{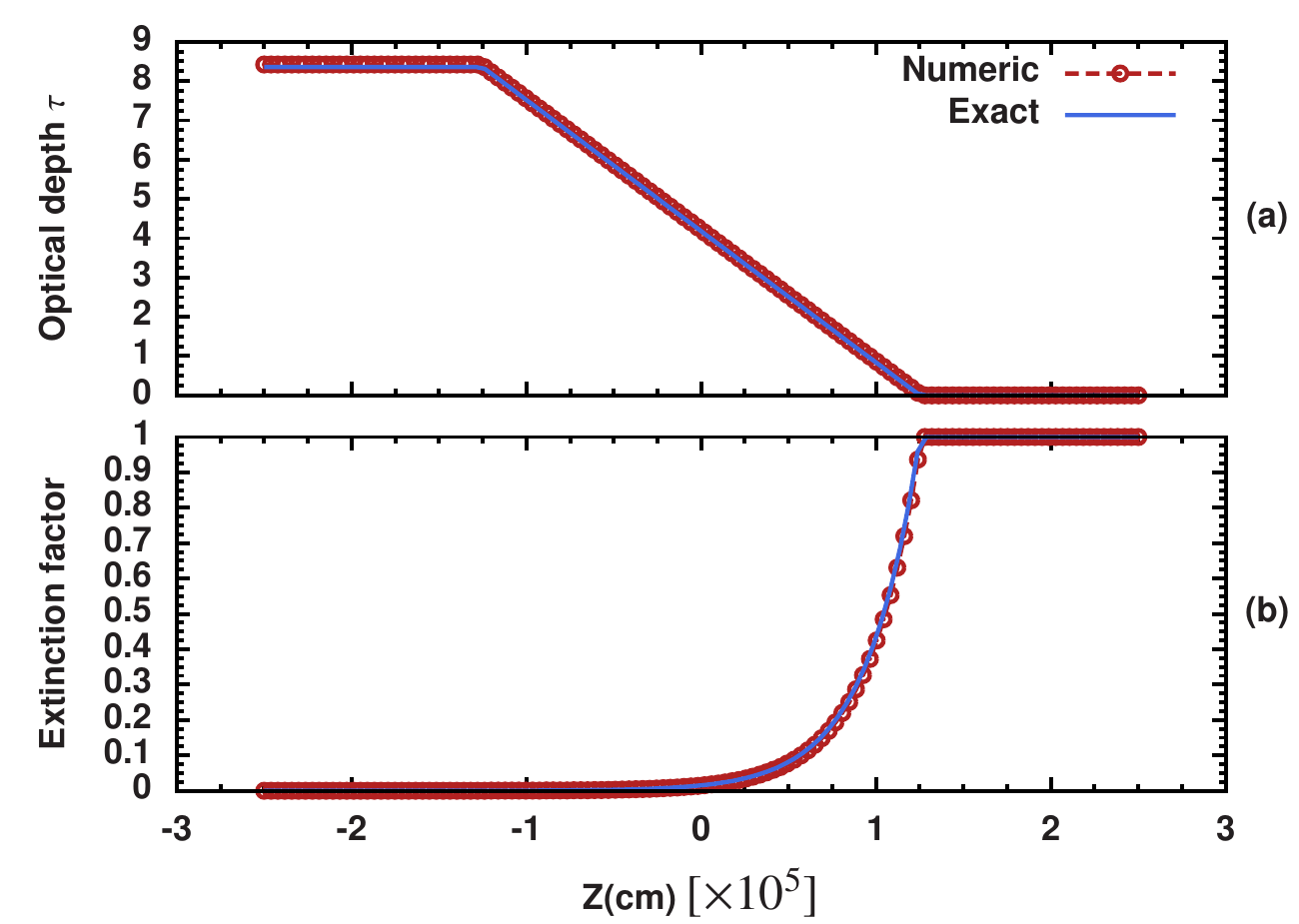}
  \caption{Optically        thin         fluid        test        (see
    Sec.~\ref{sec:rectangular-cuboid}). The upper  panel \bfa~ shows a
    comparison between  numerical and exact solutions  for the optical
    depth  in the  $z$ direction  crossing the  origin. The  detection
    surface  (observed) is  located at  $Z=2.5\times 10  ^5$ cm.   The
    difference is not noticeable in  this plot.  The lower panel \bfb~
    shows the corresponding extinction factor.}\label{fig:2}
\end{figure}

\begin{figure}
  \centering
  \includegraphics[width=84mm]{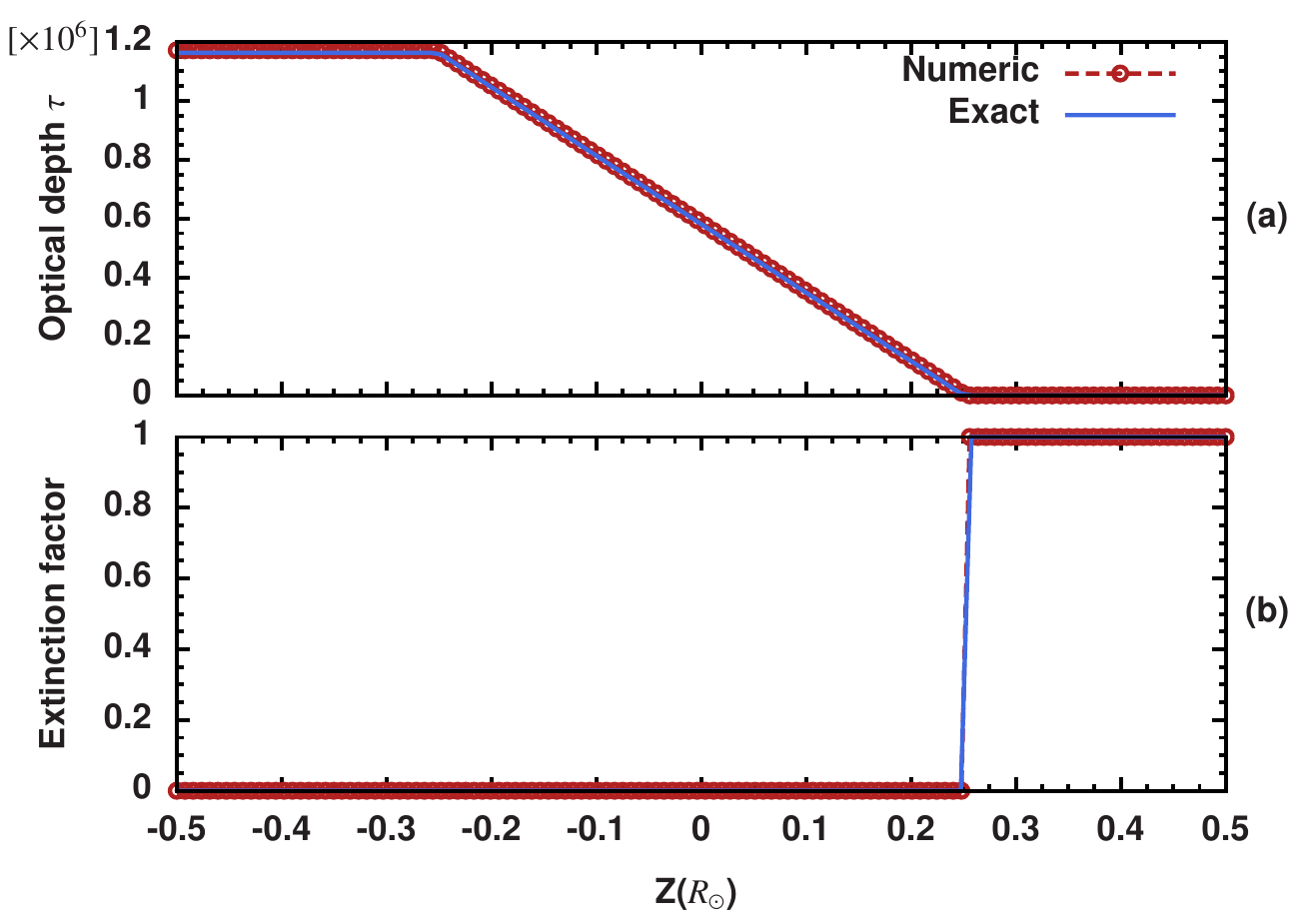}
  \caption{Optically        thick        fluid        test        (see
    Sec.~\ref{sec:rectangular-cuboid-1}). Similar to Fig. \ref{fig:2},
    the  upper panel \bfa~  shows a  comparison between  numerical and
    exact solutions for the optical depth in the $z$ direction crossing the
    origin.  The  detection surface is located  at $z=0.5\; \Rsun$.
    The difference  is not noticeable  in this plot.  The  lower panel
    \bfb~ shows the corresponding extinction factor. }\label{fig:14}
\end{figure}

Fig. \ref{fig:2}  shows a test case  for an optically  thin fluid.  
The location of the $\tau=2/3$
surface is well resolved for  the test combination of opacity, density
and  grid  size.   Similarly,  the  extinction  factor  has  a  smooth
transition between  the transparent region  ($e^{-\tau} = 1$)  and the
opaque  one ($e^{-\tau} =  0$).  On  the other  hand, Fig.~\ref{fig:14}
shows  the  numerical  solution  for  an optically  thick  fluid  (see
Section~\ref{sec:rectangular-cuboid-1}).  The  only difference between
these two  tests is the scale of  the y-axis.  In the  latter case the
$\tau=2/3$  surface is not  properly resolved  (although the  plot looks
similar,  the  y-scale  is  six  orders  of  magnitude  larger).   The
exponential factor  for the optically  thick case is a  step function:
the fluid goes  from transparent to opaque from one  grid point to the
next.

Let us consider  a cuboid $\Delta  x \times  \Delta y \times  \Delta z$
with  a fluid  at  constant temperature  $T_0$,  density $\rho_0$  and
opacity $\kappa_0$. The specific intensity in the $z$ direction is:
\begin{align}
I(\nu;T_0)&= \frac{1}{\Delta l(x,y)} \int_{-\Delta z /2}^{\Delta z/2}  B^* \e^{-\kappa_0\rho_0(\Delta z /2 - z)} \rd z \label{eq:17} \\
&= \frac{B^*}{\rho_0 \kappa_0 \Delta z}\left( 1 - \e^{-\kappa_0\rho_0\Delta z} \right) , \label{eq:18}
\end{align}
where $\Delta$l is the total length of the domain and $B^*$ is the surface brightness. The flux density then is: 
\begin{align}
\mathcal{F}(x,y)=& \frac{\sigma T_0^4}{\rho_0 \kappa_0 \Delta z} (1-\e^{-\kappa_0\rho_0\Delta z }), \label{eq:19}
\end{align}
and the luminosity in the $z$ direction is:
\begin{equation}
L_z= \Delta x \Delta y \frac{\sigma T_0^4}{\rho_0 \kappa_0 \Delta z} (1-\e^{-\kappa_0\rho_0\Delta z }). \label{eq:20}
\end{equation}
Similarly, we obtain the luminosity for the $x$ and $y$ directions.

\begin{figure}
  \centering
  \includegraphics[width=85mm]{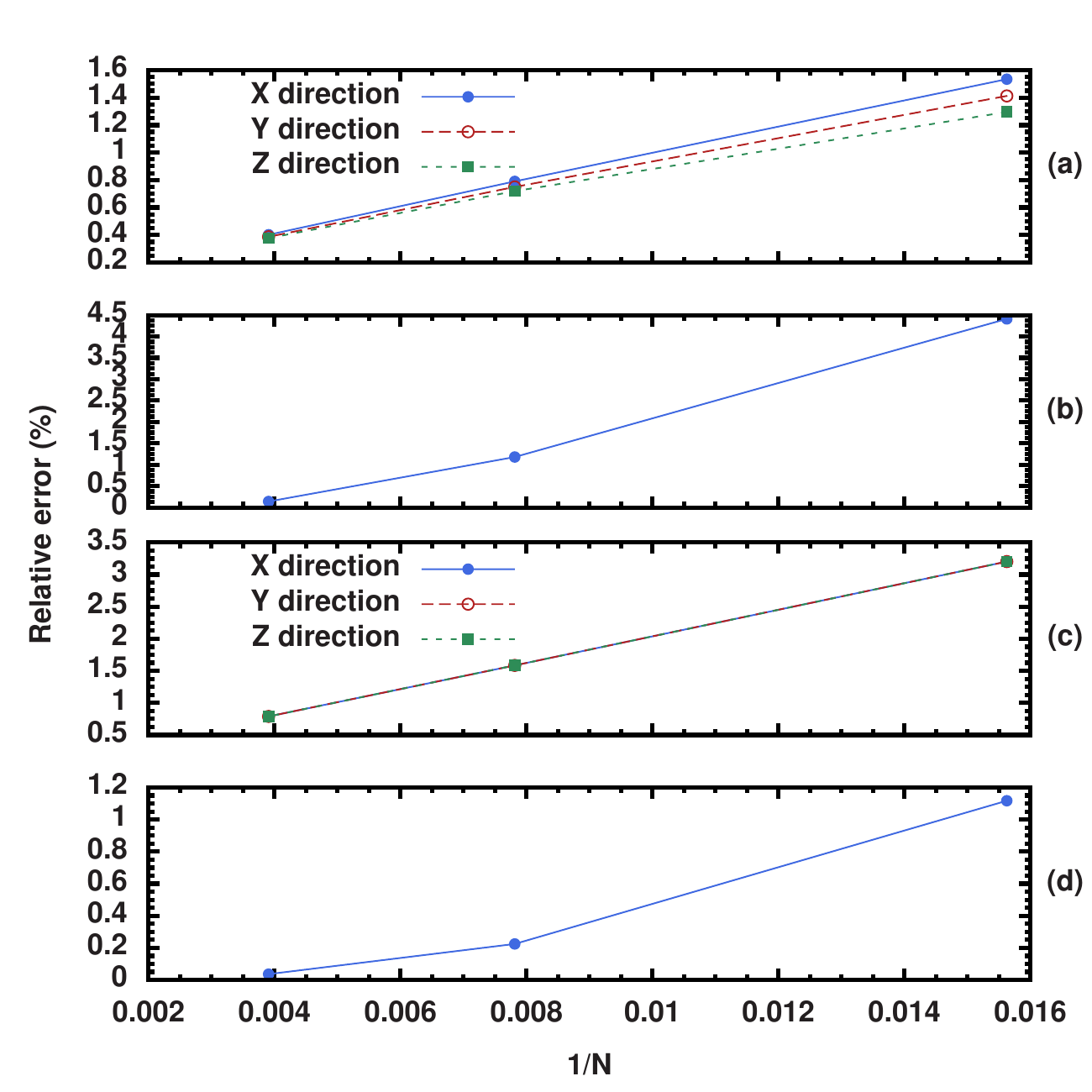}
  \caption{Convergence test. Relative error in the luminosity $\vert L
    - L_N \vert /  L$ as function of $1/N$ for grid  size $N\in \{ 64,
    128, 256 \}$  (notice that in every test  $1/N$ is proportional to
    the characteristic  gird resolution  $h$).  Panel \bfa~  shows the
    convergence for test I, panel \bfb~ shows the convergence for test
    II, panel \bfc~ shows the convergence for test III and panel \bfd~
    shows the convergence for test IV.}\label{fig:10}
\end{figure}

Using the following parameters:
\begin{align}
\kappa_0=&0.3341\quad \mathrm{cm^2/gr},\label{eq:21}\\
\rho_0=&10^{-4}\quad \mathrm{gr/cm^3},\label{eq:22}\\
T_0=&10^7 \quad \mathrm{K},\label{eq:23}\\
\Delta x = &3\times 10^5 \quad \mathrm{cm},\label{eq:24}\\ 
\Delta y = &4\times 10^5 \quad \mathrm{cm},\label{eq:25}\\ 
\Delta z = &5\times 10^5 \quad \mathrm{cm}, \label{eq:26}
\end{align}
we compare the analytic and numerical results. 
We have  already shown in  Fig.~\ref{fig:2} the optical depth  and the
extinction    factor   together    with    the   analytical    curves.
Fig.~\ref{fig:10}-\bfa\, shows  the relative error in  the luminosity as
function of resolution for  three observation directions. Note that,
since  the size  of the  cuboid is  different in  each direction,  the
corresponding  luminosity  changes  depending on  the  direction.   We
employed $1/N$ as  convergence parameter which is  proportional to the
characteristic  grid   resolution  $h$.   The  plot   shows  a  linear
dependency  $\mathcal{O}(h)$ convergence.  The relative  error in  the
luminosity is smaller than 1.6\% for each grid size.

\subsubsection{Test II. A sphere under optically thin conditions} \label{sec:test-ii-thin}

In the  case of  the cuboid the  flux density does  not depend  on the
normal  coordinates.  However,  that is  not  the case  for a  sphere.
Let's  consider a  static  sphere of  radius  $R_0$, temperature  $T_0$,
constant   opacity   $\kappa_0$   and   constant   density   $\rho_0$.
Considering  a   volume  element  in  cylindrical   coordinates.   The
parametric equation of a sphere in cylindric coordinates $\{r,\theta,z\}$ is:
\begin{align}
&\vec{r}= r \left[\cos (\theta) \hat{x}+\sin(\theta) \hat{y}\right] + z \hat{z}, \label{eq:27}\\ 
&0 \leq r \leq \sqrt{R^2-z^2}\qquad \vert z \vert \leq R. \label{eq:28} 
\end{align}
Where $R$ is the radius of the sphere.
The specific intensity in $z$ direction is 
\begin{align}
I(\nu,r;T_0)&= \frac{B^* }{2\sqrt{R^2-r^2}} \int_{-\sqrt{R^2-r^2}}^{\sqrt{R^2-r^2}}  \e^{-\kappa_0\rho_0(\sqrt{R^2-r^2} - z)} \rd z \label{eq:29}\\
&= \frac{B_{\mathrm{pk}}}{2 \rho_0 \kappa_0 \sqrt{R^2-r^2}} (1-\e^{-2 \kappa_0\rho_0 \sqrt{R^2-r^2} }), \label{eq:30}
\end{align}
the flux density is 
\begin{align}
\mathcal{F}(x,y)=& \frac{\sigma T_0^4}{2 \rho_0 \kappa_0 \sqrt{R^2-r^2}} (1-\e^{-2\kappa_0\rho_0\sqrt{R^2-r^2} }),\label{eq:31}
\end{align}
and the luminosity is 
\begin{align}
L=& 2\pi\frac{\sigma T_0^4}{\rho_0 \kappa_0} \int_0^R \frac{1-\e^{-2\kappa_0\rho_0\sqrt{R^2-r^2} }}{2\sqrt{R^2-r^2}} r \rd r, \label{eq:32} \\
=&\pi \sigma T_0^4 \frac{\e^{-2\kappa_0 \rho_0 R}+2\kappa_0\rho_0 R-1}{2\kappa_0^2\rho_0^2}. \label{eq:33}
\end{align}

Using the same parameters as above  and $R_0 = 1.25\times 10^5$~cm, we
compare  the analytic  and numerical  results (Fig.~\ref{fig:10}-\bfb).
We observe a good correspondence  between the numerical result and the
analytic one. In this case, the relative error is smaller than 4.5\%

\subsubsection{Test III. Rectangular cuboid under optically thick conditions} \label{sec:rectangular-cuboid-1}

The   difference   between   the    test   presented   here   and   in
Secs.~\ref{sec:test-iv.-sphere}   and   \ref{sec:star-model}  is   the
physical scale.  A  change in the scale induces a  situation where the
photosphere is not resolved and  requires an approximation in order to
overcome this lack of resolution.

This    test     is    similar    to    the     one    presented    in
\S~\ref{sec:rectangular-cuboid}. The  only change  is the size  of the
box,  which is  now $\Delta  x =  2$~\rsun, $\Delta  y =  1~\Rsun$ and
$\Delta z = 0.5~\Rsun$.  For  an optically thick cuboid the luminosity
in the $z$ direction is
\begin{equation}
L_z= \Delta x \Delta y \sigma T_0^4 \label{eq:34}
\end{equation}
Similarly, we obtain the luminosity for $x$ and $y$ directions.

Fig.~\ref{fig:14}  shows the  optical depth  and the  extinction factor
together with the  analytical curves. The convergence  is presented in
Fig.~\ref{fig:10}-\bfc.   Note  that in  this  case  the error  depends
exclusively on the grid size, independently of the direction. However,
the  relative  errors  are  larger   when  comparing  to  Test  I.  In
particular, the relative error is in the range [0.5\%, 3.5\%].

\subsubsection{Test IV. A sphere under optically thick conditions} \label{sec:test-iv.-sphere}

This is  the equivalent to test  II (Sec.~\ref{sec:test-ii-thin}), but
with  a  much  larger  size of  the  sphere:  $R_0=0.25\,\Rsun$.   The
luminosity radiated  by half  a solid  sphere of  constant temperature
$T_0$   and    radius   $R_0$    is   $L=2\pi    R_0^2\sigma   T_0^4$.
Fig.~\ref{fig:10}-\bfd~ shows the  corresponding convergence test.  The
convergence  in  this  case  deviates from  linear  convergence.   The
relative  errors are  the smallest  when comparing  with the  previous
test. In Section \ref{sec:star-model} we  will show that for this case
the error depends quadratically on the grid resolution.

\subsubsection{Stellar models} \label{sec:star-model}

Although it  is possible  to compute the  luminosity for  an arbitrary
fluid  distribution, \Celmo  ~is  designed to  work with  hydrodynamic
evolution  of stars.  Here  we  compute the  luminosity  of two  stars
calculated  by  the  1D  code  \Mesa\  and  mapped  into
\Enzo\ using 128$^3$ and 256$^3$ cell resolutions.

We use  two \Mesa~ profiles  in order to  test the computation  of the
initial luminosity, a smaller red-giant-branch (RGB) star and a larger
asymptotic-giant-branch  (AGB) star  whose  parameters  are listed  in
Table \ref{tab:1}.   Both of these  stars are in the  optically thick
regime.  Therefore,  the results  are similar  to the  optically thick
solid sphere (Sec.  \ref{sec:test-iv.-sphere}).   The stars cover most
of the numerical  domain.  In Table ~\ref{tab:1} we see  how the error
is reduced with  increasing resolution.  This serves  as an additional
verification  tests and  gives a  measure  of the  uncertainty on  the
luminosity  when there  is  no uncertainty  in  temperature.  This  is
effectively dominated by locating the  photosphere which is in turn is
uncertain by at most a resolution element.

The approximate error in the calculation of the luminosity is given by
the errors on the temperature ($\Delta  T$) and on the radius ($\Delta
R_*$):
\begin{equation}
\Delta L = \frac{\partial L}{\partial R}\Delta R + \frac{\partial L}{\partial T} \Delta T \label{eq:35}
\end{equation}
For  a  star  of  radius $R_*$,  effective  temperature  $T_\eff$  and
luminosity $L_* = 4 \pi \sigma R_*^2 T_\eff^4$, the relative error is:
\begin{equation}
\frac{\Delta L_*}{L_*} = 2 \frac{\Delta R}{R_*} + 4\frac{\Delta T}{T_\eff} \label{eq:36}
\end{equation}

For a  numeric grid with  $\Delta x$, $\Delta  y$ and $\Delta  z$, the
radius $R_{\rm phot}$ of our initial model is located in a cell defined
by  $(x_i,y_j,z_k)$ and  $(x_{i+1},y_{j+1},z_{k+1})$.  Therefore,  the
difference between  the radius of  the star  and the numerical  one is
$\Delta R  = \vert R_*  - R_{ijk} \vert  $ and satisfy  the inequality
$\Delta R \le \sqrt{ (\Delta x)^2  + (\Delta y)^2 + (\Delta z)^2}$. On
the  other hand,  the $\Delta  T =  0$ since  we initially  assign the
effective   temperature   $T_\eff$  to   the   grid   points  in   the
photosphere. The  estimation of  the errors  in Table  \ref{tab:1} are
calculated under these assumptions.

\begin{table*}
\centering
\begin{minipage}{126mm}
  \centering
  \caption{Parameters of two \Mesa~ models and the luminosity computed 
    in \Celmo~ ($L_{\Celmo}$) for two resolutions.  
    The measurement uncertainty in the luminosity is estimated using 
    Equation~\eqref{eq:36}.}\label{tab:1}
\begin{tabular}{cccccc}
    \hline  \hline
       profile &    $R\; (\Rsun)$ &  $T_{\eff}$ (K) & $L\; (\Lsun)$  & \multicolumn{2}{c}{$L_\Celmo\; (\Lsun)$} \\
       &&&& $128^3$ & $256^3$\\ \hline
       RGB & 83 & 3200 & 648 & 662 $\pm$ 94 & 638 $\pm$ 46\\
       AGB &  170   & 3042 & 2207  & 2202 $\pm$ 154 &  2204 $\pm$ 76\\
    \hline
\end{tabular}
\end{minipage}
\end{table*}

\newpage
\bibliographystyle{mn2e}
\bibliography{bibliography}


\end{document}